\documentclass[10pt,journal,compsoc,twoside]{IEEEtran}
\usepackage{epsfig,endnotes,multirow,adjustbox,amsmath,mathtools,todonotes,hyperref,fancyhdr,paralist,xcolor,booktabs,svg}%
\usepackage{support-caption}
\usepackage{subcaption}

\ifCLASSOPTIONcompsoc
  \usepackage[nocompress]{cite}
\else
  \usepackage{cite}
\fi
\ifCLASSINFOpdf
\else
\fi
\begin{document}
%
%
\title{On the Vulnerability of Data Points under Multiple Membership Inference Attacks and Target Models}

%
%
%
%

\author{Mauro~Conti,~\IEEEmembership{Fellow,~IEEE,}
        Jiaxin~Li*,
        and~Stjepan~Picek,~\IEEEmembership{Senior~Member,~IEEE}
\IEEEcompsocitemizethanks{\IEEEcompsocthanksitem Mauro Conti and Jiaxin Li (corresponding author) are with the Department of Mathematics, University of Padua,
35131 Padua, Italy (Email: conti@math.unipd.it, jiaxin.li@studenti.unipd.it). Mauro Conti is also affiliated with TU Delft and the University of Washington.\protect\\
\IEEEcompsocthanksitem Stjepan Picek is with the Institute for Computing and Information Sciences, Radboud University, 6525 EC Nijmegen, Netherlands (Email: stjepan.picek@ru.nl
).\protect\\
\IEEEcompsocthanksitem * means the corresponding author.}}
\IEEEtitleabstractindextext{%
\begin{abstract}
Membership Inference Attacks (MIAs) infer whether a data point is in the training data of a machine learning model. It is a threat while being in the training data is private information of a data point. MIA correctly infers some data points as members or non-members of the training data. Intuitively, data points that MIA accurately detects are vulnerable. Considering those data points may exist in different target models susceptible to multiple MIAs, the vulnerability of data points under multiple MIAs and target models is worth exploring.\\

This paper defines new metrics that can reflect the actual situation of data points' vulnerability and capture vulnerable data points under multiple MIAs and target models. From the analysis, MIA has an inference tendency to some data points despite a low overall inference performance. Additionally, we implement 54 MIAs, whose average attack accuracy ranges from 0.5 to 0.9, to support our analysis with our scalable and flexible platform, Membership Inference Attacks Platform (VMIAP). Furthermore, previous methods are unsuitable for finding vulnerable data points under multiple MIAs and different target models. Finally, we observe that the vulnerability is not characteristic of the data point but related to the MIA and target model.
\end{abstract}

\begin{IEEEkeywords}
Machine learning, privacy, membership inference attack, vulnerable data points.
\end{IEEEkeywords}}

\maketitle

\IEEEdisplaynontitleabstractindextext

%
\IEEEpeerreviewmaketitle


\IEEEraisesectionheading{\section{Introduction}
\label{sec:introduction}}

\IEEEPARstart{M}achine learning, especially with the development of deep learning, promotes many real-world applications, e.g., computer vision, natural language processing, and data mining. With machine learning's frequent practical applications, security and privacy problems are exposed, including models' fairness~\cite{Mehrabi_2021_fairness,Chang_2021_Fairness}, adversarial examples~\cite{Goodfellow_2015_adversarial,Yuan_2019_adversarial}, and parameters stealing~\cite{Florian_2016_steal,Wang_2018_steal}. A membership inference attack (MIA) detects whether a data point is in the training data of a machine learning model by which it violates the data points' privacy. MIA became an important topic after the seminal work by Shokri et al.~\cite{shokri_membership_2017}. A successful MIA has serious consequences, especially when training data is sensitive or private, like medical data, bank account information, or historical browsing records. With the release of laws about private data usage, like the General Data Protection Regulation (GDPR) in Europe, and the California Consumer Privacy Act (CCPA) in the United States, more attention is given to the security and privacy of data. Consequently, MIAs have become one of the focus points for academia and industry.

The factors connected with implementing MIAs (and their success) are diverse. Some papers mention possible reasons for the success of MIA. It is reported that different model structure choices, datasets of training models~\cite{truex_towards_2018}, and subgroups of the dataset for evaluation ~\cite{yaghini_disparate_2019} influence MIA's success. Notably, overfitting is the most frequently mentioned reason. Yeom et al. showed that overfitting is a sufficient but unnecessary condition for MIA~\cite{yeom_privacy_2018}. Moreover, Long et al. showed a pragmatic MIA to well-generalized models, which also means overfitting is not the necessary reason~\cite{long_pragmatic_2020}. Yaghini et al.~\cite{yaghini_disparate_2019}, and Da et al.~\cite{Da_2022_Understanding} found that a subgroup of data points with one or several sharing attributes is more vulnerable to MIA. Generally, several elements contribute to the success of MIAs, which provide the baseline for designing and analyzing our experiments.

It has been empirically demonstrated that some data points in the dataset are more vulnerable to MIA~\cite{long_pragmatic_2020,song_systematic_2021,duddu_shapr_2021}. In general, vulnerable data points are those data points correctly inferred with a high probability. Some research works propose various approaches to determine specific vulnerable data points. Long et al. selected data points with fewer neighbors on the combination feature space represented by outputs of some reference models as vulnerable ones~\cite{long_pragmatic_2020}. We denote this method as the neighbors-based method in our paper.
Furthermore, the authors mentioned that finding vulnerable data points with the outlier detection method is possible. Therefore, we leverage one outlier detection method (SUOD)~\cite{zhao_suod_2020} to detect possible vulnerable data points, which provides a comparison choice. In 2021, Song et al. defined the privacy risk score of a single data point as its probability of being a member under the condition that the target model's output on this data point is observed~\cite{song_systematic_2021}. After that, Duddu et al. proposed to use the Shapley value to measure data points' susceptibility to MIAs~\cite{duddu_shapr_2021}. However, those methods either do not analyze each data point separately or do not explore the data point's vulnerability under multiple MIAs and different target models. Besides, they mainly focus on the data point's vulnerability while the data point is in the training data and do not explore the case that the data point is in the testing data. Our paper explores vulnerable data points under multiple MIAs and different target models from empirical results to rectify those aspects. Then, we compare and analyze vulnerable data points we find with those found by previous methods.

We explore the vulnerability of data points under multiple MIAs and target models. It is worth noting that those target models are trained with different data splits of the same dataset and have the same size of data points, identical model structure, training process, and hyperparameters while analyzing a single data point's vulnerability. 
As we mentioned before, MIA is related to diverse factors. Those target models trained with different data splits of the same dataset will avoid the possible difference in attack performance brought by the difference in sub-dataset size, model structure, training process, and hyperparameters of target models. In addition, we define several metrics about data points' exposure rate and MIA's inference rate, as discussed in Section~\ref{definitions} to formally analyze data points' vulnerability to multiple MIAs and different target models. The main reason for new definitions is that previous research works~\cite{long_pragmatic_2020,song_systematic_2021,duddu_shapr_2021} on the vulnerable data points do not provide suitable metrics to evaluate the data point's privacy risk. 
Moreover, to make it convenient and fast to attack target models with multiple MIAs, we design and implement the VMIAP\footnote{We will release the code after the paper is accepted.}, a scalable and flexible platform to conduct various MIAs against target models. Finally, we compare vulnerable data points determined by the empirical result with previous works. 

Our main contributions are:
\begin{compactenum}
    \item We define metrics related to the data point's exposure rate and MIA's inference rate to analyze data points' vulnerability. Observing the results of experiments, we find that two new metrics can reflect the actual situation of data points' vulnerability and capture vulnerable data points under multiple MIAs and target models. From analysis, there is indeed a part of vulnerable data points under multiple MIAs and target models. Furthermore, if the data point appears in the training data of target models, its vulnerability is usually higher than in the testing data. Finally, we show that MIA has an inference tendency to some data points despite overall low inference performance.

    \item We design and implement a scalable and flexible platform VMIAP for attacking different target models with multiple types of MIAs and their variants. The average attack accuracy of MIAs ranges from 0.5 to 0.9. We observe that the accuracy gap, dataset, and model structure affect attack performance. 

    \item We analyze and compare vulnerable data points determined by our new metrics and previous methods. We show those previous methods are unsuitable for detecting vulnerable data points under multiple MIAs and target models. Besides, we find that the vulnerability is not the fixed characteristic of the data point but related to the MIA and target model.

\end{compactenum}

\section{Background and Threat Models}

We present the necessary background information about machine learning and the Membership Inference Attack Game in Section~\ref{machine_learning_background} and Section~\ref{MIA_game}. Then we discuss the threat models of MIAs implemented in our paper.

\subsection{Machine Learning}
\label{machine_learning_background}

The target model we focus on is a supervised machine learning model for classification tasks. The training dataset $D$, drawn from the underlying distribution \textbf{D}, is composed of a large number of data points $(x_i,y_i)$. The $x_i$ and $y_i$ represent the data point's feature vector and label. The unique values of $y_i$ indicate categories of data points. The neural network model $f_{\theta}$ obtains the feature vector $x_i$ and predicts its label based on the maximum probability it belongs to each category. The training of the $f_{\theta}$ model makes the outputs of the model to data points in the training dataset close to their ground truths. The learning algorithm and loss function guide the adjustment of the model's parameters $\theta$. The iterative update of $\theta$ is along a gradient descent direction to minimize the loss of a batch of data points in the training dataset. The model will reuse data points of the training dataset for updating parameters several times. After the training of $f_{\theta}$, the model will evaluate the testing dataset, which has no overlapping data points with the training dataset. The prediction accuracy of $f_{\theta}$ on the testing dataset presents the model's generalization ability. The model with a high prediction accuracy on the testing dataset is more effective in practice. 

\subsection{Membership Inference Attack Game} 
\label{MIA_game}
We use the definition of Membership Inference Attack Game presented by Yeom et al.~\cite{yeom_privacy_2018} and followed in the works of Carlini et al.~\cite{carlini_membership_2021}, and Jayaraman et al.~\cite{Jayaraman_2021_revisiting}. 

\textbf{Definition 1} (Membership Inference Attack Game)\textbf{.} \textit{The game between a challenger C and an adversary A:}

1) \textit{The challenger samples a training dataset $D$ from the underlying distribution \textbf{D} and trains a model $f_\theta$ based on $D$.}

2) \textit{The challenger flips a bit $b$. If $b = 0$, the challenger samples a data point $(x,y)$ from the distribution \textbf{D}. If $b = 1$, the challenger randomly selects a data point $(x,y)$ from the training dataset $D$.}

3) \textit{The challenger sends $(x,y)$ to the adversary.}

4) \textit{The adversary queries the target model $f_\theta$ and has some extra knowledge $E$. Then, the adversary outputs the prediction $b'=A_{f_\theta,E,I}(x,y)$ with specific utilization strategy $I$ of knowledge.}


5) \textit{Outputs 1 if $b = b'$, else 0.}

If \textit{$b = b'$}, the game's output is 1, which means the adversary wins this game on a single data point. Otherwise, the challenger successfully defends and makes the prediction of the adversary wrong. In realistic experiments, we act as both the challenger and the adversary to test the performance of adversaries or MIAs. The test or evaluation depends on the result of a specific adversary on a large number of data points.





\subsection{Threat Model}
\label{threat_model}

As the definition of Membership Inference Attack Game in the previous section, the adversary has access to the target model \textit{$f_\theta$} and some extra knowledge \textit{$E$}. The adversary's purpose is to win the Membership Inference Attack Game on data points as many times as possible. If the adversary obtains different information \textit{$E$} and changes the way {$I$} of utilizing information (or messages), we regard it as a new type of MIA in our paper. We implement 54 different types of MIA to explore data points' vulnerability, which is the focus of this research. The basic ideas of those MIAs follow previous works. Then, we change the details of MIAs' implementations and construct variants to compare MIAs under the same environment and extend possible related MIAs to evaluate our newly defined metrics. To formally explain each MIA's threat model, we expose multiple MIAs' possible messages and utilization strategies in our paper. The possible messages and utilization strategies distinguish different MIAs.

\textbf{Possible messages.} An additional dataset \textit{$D'$} (called the shadow data), which is sampled from the distribution \textit{\textbf{D}} and has no overlapping data points with the target dataset $D$ (called the target data), the unique values of label \textit{y}, the type and structure of the target model \textit{$f_{\theta}$}, the training process of the target model \textit{$f_{\theta}$}, the hyperparameters while training the target model \textit{$f_{\theta}$}, and the output of the target model $f_{\theta}$ for the input $x$.

\textbf{Utilization strategies.} Utilization strategies are composed of two categories. One represents approaches to calculate metrics based on the output of \textit{$f_{\theta}$} to data point's feature vector \textit{x} and label \textit{y}. Those metrics consist of the prediction probability vector, the maximum probability value, the probability of ground truth, the cross-entropy loss, the CELoss defined by Li et al.~\cite{li_label-leaks_2020}, the Mentr value defined by Song et al.~\cite{song_systematic_2021}, the entropy of the prediction probability vector and its normalization value. The other category is the types of binary model (the attack model) for predicting \textit{$A_{f_\theta, E}(x,y)$} for the adversary. We select Support Vector Machine (SVM)~\cite{noble_what_2006}, Linear Model~\cite{shao_linear_1993}, XgBoost~\cite{chen_xgboost_2016}, and shallow MultiLayer Perceptron (MLP)~\cite{ramchoun_multilayer_2016} as possible architectures of the attack model. Including the classical model, ensemble model, and neural network brings diversity to MIAs, making comparison and evaluation convenient. Apart from the attack model, the adversary can also compare calculated metrics with thresholds for attacking. This means there are two ways of implementing MIA: the attack model and threshold comparison.

The basic processes of implementing MIAs in both methods include:
\begin{compactenum}
\item Train the shadow model with the shadow data.
\item Extract shadow features or metrics by feeding the shadow data to the shadow model.
\item Train the attack model with shadow features or determine thresholds with shadow metrics.
\item Extract target features or metrics by putting the target data into the target model.
\item Attack the target model by putting target features into the attack model or comparing target metrics with thresholds.
\end{compactenum}


\section{New Metrics for Data Point and MIA}
\label{definitions}

Previous works~\cite{long_pragmatic_2020,song_systematic_2021,duddu_shapr_2021} about the vulnerable data point, the data point's privacy risk score, and the Shapley value are unsuitable for describing vulnerable data points under multiple MIAs and target models. Indeed, those methods either do not analyze each data point separately or do not explore the data point's vulnerability under multiple MIAs and target models. Therefore, we define several metrics about the data point's exposure rate and MIA's inference rate. Section~\ref{exposure_rate} explains some metrics of the data point's exposure rate. Section~\ref{inference_rate} interprets some metrics of MIA's inference rate. Then, we determine vulnerable data points under multiple MIAs and target models with the help of two newly defined metrics. Here, we consider that different target models are trained with varying data splits of the same dataset and have the same training process, hyperparameters, and model structure. As we mentioned in Section~\ref{threat_model}, if the adversary obtains different messages and changes how they utilize them, we regard it as a new type of MIA.

\subsection{Data Point's Exposure Rate}
\label{exposure_rate}

A single data point in the target data of a target model might be in the training data, which means member, or in the testing data, which implies non-member. There are several MIAs that can attack this target model. Thus, we define the data point's exposure rate, which represents this data point's frequency of being correctly inferred by those MIAs while this data point is in the target data of the target model. Besides, we distinguish the situation in which this data point is in the training or testing data, meaning member and non-member exposure rates separately. Considering this data point may appear in the target data of different target models, we count the number of target models whose training data includes this data point as its Member Times (MT). In addition, the number of target models whose testing data consists of this data point is the Non-Member Times (NMT) of this data point. Those different target models are models that have the same training process, hyperparameters, and model structure. The only difference is that we train those target models from various splits of the same dataset.
\textbf{Definition 2} (Data point's Member Exposure Rate)
Within different dataset splits, different target models are trained with identical model structure, training process, and hyperparameters. The training data size of each target model is the same and marked as \textit{$|D|$}. A specific data point \textit{(x,y)} is in the training data of \textit{m} target models, which means \textit{b(x,y)=1} before sending this data point to the adversary in the Membership Inference Attack Game. The \textit{m} is the MT of this data point. For each target model \textit{$f_{(\theta,j)}$} in those \textit{m} target models, there are \textit{n} membership inference attacks targeting each target model. The \textit{i}-th adversary is denoted as \textit{$A_{(f_{(\theta,j)},E_i,I_i)}$}. In this notation of the adversary, the \textit{$E_i$} is the additional data for the \textit{i}-th adversary, \textit{$f_{(\theta,j)}$} is the \textit{j}-th target model, \textit{$I_i$} indicates the message utilization strategies of the \textit{i}-th adversary. The output of the adversary to this data point \textit{(x,y)} is \textit{$b^{'}$}, which means \textit{$b^{'}_{(j,i)} = A_{(f_{(\theta,j)},E_i,I_i)}(x,y)$}. The Member Exposure Rate (MER) of this data point \textit{(x,y)} under the \textit{j}-th target model is defined as follows:

\begin{equation}
\label{MER_formula}
\textit{$MER_{j}(x,y)=\frac{\sum_{i=1}^{n} B(b^{'}_{(j,i)}(x,y) == b(x,y))}{n}$}.
\end{equation}
Here, the \textit{b(x,y)} is always equal to 1 because the data point \textit{(x,y)} is in the training data of the target model. The \textit{B} is the indicator function, which means \textit{B(true)=1} and \textit{B(false)=0}. This indicator function \textit{B} used in other definitions has the same meaning. The \textit{n} is the number of MIAs against the target model. The member exposure rate shows how frequently those MIAs correctly infer the data point in the training data of one target model. 

To further describe the data point's vulnerability under \textit{m} target models, we define the Average Member Exposure Rate (AMER) of the data point \textit{(x,y)} as:
\begin{equation}
\label{AMER_formula}
\textit{$AMER(x,y)=\frac{\sum_{j=1}^{m} MER_{j}(x,y)}{m}$}.
\end{equation}
In the above expression, the \textit{m} is the number of target models whose training data includes the data point \textit{(x,y)}, i.e., the MT of this data point.
 
 

\textbf{Definition 3} (Data point's Non-Member Exposure Rate) The data point's Non-Member Exposure Rate (NMER) definition is similar to MER. The only difference is that the data point \textit{(x,y)} is in the testing data of different target models, which means \textit{b(x,y)=0} before sending the data point to the adversary. The number of different target models whose testing data includes the data point \textit{(x,y)} is $k$. The NMER of the data point \textit{(x,y)} on the \textit{j}-th target model is defined as:
\begin{equation}
\label{NMER_formula}
\textit{$NMER_{j}(x,y)=\frac{\sum_{i=1}^{n} B(b^{'}_{(j,i)}(x,y) == b(x,y))}{n}$}.
\end{equation}
Here, \textit{b(x,y)} is always equal to 0 because the data point \textit{(x,y)} is in the testing data of the target model. The \textit{n} is the number of MIAs. Similarly, we define the Average Non-Member Exposure Rate (ANMER) of the data point \textit{(x,y)} as:
\begin{equation}
\label{ANMER_formula}
\textit{$ANMER(x,y)=\frac{\sum_{j=1}^{k} NMER_{j}(x,y)}{k}$}.
\end{equation}
In the above expression, the \textit{k} is the number of target models whose testing data includes data point \textit{(x,y)} and also means the NMT of this data point.

The definitions of AMER and ANMER measure the vulnerability of data points from the actual inference correctness of MIAs against target models. Therefore, we can determine vulnerable data points under multiple MIAs and target models with the help of those two metrics from the perspective of the empirical result. The data points with a relatively high value of AMER or ANMER are selected as vulnerable data points under multiple MIAs and target models. Besides, we can analyze the AMER value, the ANMER value, and the difference between those two metrics of different data points. The analysis compares the AMER values of data points, ANMER values of data points, and one data point's AMER and ANMER values.

The above definitions are from the perspective of the data point to detect vulnerable data points under multiple MIAs and target models. We can also analyze the inference correctness from the angle of MIA. Let us consider that a specific MIA correctly infers the existence of a single data point in a target model. Can this MIA infer the data point's presence while attacking other target models? What is the inference correctness of this MIA to other data points? Does this MIA have an inference tendency to a part of data points? To understand those questions, we define metrics about MIA's inference rate from the angle of the MIA to analyze its inference correctness to data points. 

\subsection{MIA's Inference Rate}
\label{inference_rate}

While attacking a target model, the MIA correctly infers a single data point's existence in this target model's training or testing data. Similarly, different target models are models trained from varying data splits with identical model structures, training processes, and hyperparameters. As we mentioned, the number of target models whose training data includes this single data point is this data point's MT. Furthermore, the number of target models whose testing data consists of this single data point is the NMT of this data point. The MIA's inference rate to one data point means the percentage of correct inference to this data point among target models. Considering this data point is in the training or testing data, we define the member inference or non-member inference rates.

\textbf{Definition 4} (MIA's Member Inference Rate)
The training data size in each target model is the same and marked as \textit{$|D|$}. A specific data point \textit{(x,y)} is in the training data of \textit{$m_{(x,y)}$} target models among those models, which means \textit{b(x,y)=1} before sending the data point to the adversary. The \textit{$m_{(x,y)}$} is the MT of this data point. In those \textit{$m_{(x,y)}$} target models, there are \textit{n} MIAs target at those models and the \textit{i}-th adversary is denoted as \textit{$A_{(f_{(\theta,j)},E_i,I_i)}$}, where \textit{$E_i$} is the additional data for \textit{i}-th adversary, \textit{$f_{(\theta,j)}$} is \textit{j}-th target model, and \textit{$I_i$} indicates the message utilization strategies of the \textit{i}-th adversary. The output of the adversary is \textit{$b^{'}$}, i.e., \textit{$b^{'}_{(j,i)} = A_{(f_{(\theta,j)},E_i,I_i)}(x,y)$}. The Member Inference Rate (MIR) of \textit{i}-th MIA to the data point \textit{(x,y)} under \textit{$m_{(x,y)}$} target models is defined as following:


\begin{equation}\label{MIR_formula}
\textit{$MIR_{i}(x,y)=\frac{\sum_{j=1}^{m_{(x,y)}} B(b^{'}_{(j,i)}(x,y) == b(x,y))}{m_{(x,y)}}$}.
\end{equation}
Here, \textit{b(x,y)} is always equal to 1 because the data point \textit{(x,y)} is from the training data of the target model. The \textit{$m_{(x,y)}$} is the number of target models whose training data includes the data point \textit{(x,y)}. Among different splits of the dataset \textit{D} for training target model, the number of data points appearing in the training data of target models is Member Number (MN). To further describe MIA's inference correctness to other data points, we define the Average Member Inference Rate (AMIR) of the \textit{i}-th MIA as:
\begin{equation}\label{AMIR_formula}
\textit{$AMIR_{i}(x,y)=\frac{\sum_{k=1}^{MN} MIR_{i}(x_{k},y_{k})}{MN}$}. 
\end{equation}
\textit{MN} is the number of data points that appear in the training data of target models and \textit{$(x_{k},y_{k})$} is the \textit{k}-th data point among them. Each of data point might has different MT, represented as \textit{$m_{(x_{k},y_{k})}$}.

\textbf{Definition 5} (MIA's Non-Member Inference Rate) The definition of the data point's Non-Member Inference Rate (NMIR) is similar to the MIR. The only difference is that the data point \textit{(x,y)} is in the testing data of target models, i.e., \textit{b(x,y)=0} before sending the data point to the adversary. The number of target models whose testing data includes the data point \textit{(x,y)} is \textit{$z_{(x,y)}$}. In other aspects, the NMIR is similar to the MIR. The NMIR of \textit{i}-th MIA to the data point \textit{(x,y)} under \textit{$z_{(x,y)}$} target models is defined as:
\begin{equation}\label{NMIR_formula}
\textit{$NMIR_{i}(x,y)=\frac{\sum_{j=1}^{z_{(x,y)}} B(b^{'}_{(j,i)}(x,y) == b(x,y))}{z_{(x,y)}}$}.
\end{equation}
Here, \textit{b(x,y)} is always equal to 0 because the data point \textit{(x,y)} is from the testing data of the target model. The \textit{$z_{(x,y)}$} is the number of target models whose testing data includes the data point \textit{(x,y)}. Among different splits of the dataset \textit{D} for training target models, the number of data points appearing in the testing data of target models is Non-Member Number (NMN). Similarly, we define the Average Non-Member Inference Rate (ANMIR) of the MIA \textit{i} as:
\begin{equation}\label{ANMIR_formula}
\textit{$ANMIR_{i}(x,y)=\frac{\sum_{k=1}^{NMN} NMIR_{i}(x_{k},y_{k})}{NMN}$},
\end{equation}
where \textit{NMN} is the number of data points appearing in the testing data of target models and \textit{$(x_{k},y_{k})$} is the \textit{k}-th data point among them. Each of data point might have different NMT, which is represented as \textit{$z_{(x_{k},y_{k})}$}.

We can analyze MIA's MIR and NMIR values against other data points. With those metrics about MIA's inference rate, we can answer questions including MIA's inference correctness to different data points and MIA's inference tendency to a part of data points.

\subsection{Explanation of Exposure and Inference Rate}

Considering that the definitions of data points' exposure rate and MIA's inference rate are abstract, a simple example is beneficial to explain those metrics. Due to the metrics for a data point in the training data and the metrics for a data point in the testing data being similar, we illustrate the (Average) Member Inference Rate and (Average) Member Exposure Rate in Table~\ref{simple_example}. 


\begin{table*}[ht!]
\centering
\caption{(Average) Member Inference Rate and (Average) Member Exposure Rate}
\label{simple_example}
\begin{adjustbox}{max width=1\textwidth}
\begin{tabular}{|c|c|c|c|c|c|c|c|c|c|c|c|c|}
\hline
& \multicolumn{12}{c|}{Member} \\

\hline
& \multicolumn{4}{c|}{D\_1}                & MIR               & \multicolumn{5}{c|}{D\_2}                    & MIR           & AMIR                  \\

\hline
M\_1 & 1        & 0        & 0        & 1      & 2/4               & 1       & 1      & 0      & 1      & 0      & 3/5           & (2/4+3/5)/2           \\

\hline
M\_2 & 1        & 0        & 1        & 1      & 3/4               & 1       & 0      & 1      & 1      & 1      & 4/5           & (3/4+4/5)/2           \\

\hline
MER  & 2/2      & 0/2      & 1/2      & 2/2    &  & 2/2     & 1/2    & 1/2    & 2/2    & 1/2    & \multicolumn{1}{c}{}&\\

\hline
AMER & \multicolumn{4}{c|}{(2/2+0/2+1/2+2/2)/4} &                   & \multicolumn{5}{c|}{(2/2+1/2+1/2+2/2+1/2)/5} & \multicolumn{2}{c|}{}\\

\hline

\end{tabular}
\end{adjustbox}
\end{table*}

In Table~\ref{simple_example}, \textit{M\_i} means the i-th MIA, and \textit{D\_i} means the i-th data point. The number 0 represents a data point in the training data of the target model but is inferred as a non-member by the MIA. The number 1 means the data point is in the training data of the target model and is predicted as a member by the MIA, which is correct inference. \textit{D\_1} in the training data of 4 target models and \textit{D\_2} in the training data of 5 target models from observations that 4 and 5 inference values (0 or 1) between the MIA and data point. The number of MIAs is 2 (\textit{M\_1} and \textit{M\_2}), and the number of data points is 2 (\textit{D\_1} and \textit{D\_2}). Here, the number of target models, data points, and MIAs in the table are assumed to explain metrics, meaning they are not actual quantity values in our experiments. From the table, we can obtain that MIR, AMIR, MER, and AMER are calculated based on the inference values.

\section{Our Various Membership Inference Attacks Platform (VMIAP)}
\label{VMIAP}

Previous code implementations of membership inference attacks do not provide the opportunity of applying different types of MIAs to target models. Furthermore, they do not analyze a single data point's vulnerability under multiple MIAs and target models, which are not suitable for our research exploration. Thus, we designed and implemented the VMIAP, which is scalable and flexible, to support our research about vulnerable data points under multiple MIAs and target models. In Section~\ref{different_type_of_MIAs}, we explain different types of MIAs. Then, we provide the details of platform implementation in Section~\ref{platform_implementation_detail}. 

\subsection{Different Types of MIAs}
\label{different_type_of_MIAs}

The membership inference attack, proposed by Shokri et al.~\cite{shokri_membership_2017}, undergoes a series of developments. As we mentioned in Section~\ref{threat_model}, if the adversary obtains different messages and changes how to utilize them, we regard it as a new type of MIA in our paper. Most MIAs use additional data (called shadow data, reference data, and additional data in different works) from the same distribution with the training data of target models to assist with attacking. The basic processes of implementing MIA are explained in Section~\ref{threat_model}.


We categorize MIAs into two types depending on the utilization of features or metrics. One is classifier-based MIA which trains a classifier (the attack model) to distinguish a member and a non-member. The other is non-classifier-based MIA that implements MIA by comparing with the threshold. The features while training the classifier, the model architecture of the classifier and metrics for determination are factors of different MIAs. Besides, we use relabelling to deal with the situation where the prediction probability vector of the target model is not exposed, and the target model only outputs the label of the query, which means a label-only condition. The relabelling means relabelling shadow data with the target model and obtaining the features and metrics of target data from the relabelled shadow model, as in the work of Li et al.~\cite{li_label-leaks_2020}.
Additionally, the gap MIA, proposed in the work of Yeom et al.~\cite{yeom_privacy_2018}, is also included for comparison. The gap MIA classifies the data point whose prediction of the target model is equal to the ground truth as a member. Otherwise, it is classified as a non-member.

The adversaries extract features and metrics from the output of the model. The model's output is the prediction probability vector. In addition to it, the maximum probability value of the output, the probability of ground truth, the cross-entropy loss computed with Pytorch function CrossEntropyLoss, the CELoss defined in Li's work~\cite{li_label-leaks_2020}, the Mentr value defined in Song's work~\cite{song_systematic_2021}, the entropy of prediction vector~\cite{lesne_shannon_2014}, and its normalization value~\cite{shokri_membership_2017} are possible metrics for non-classifier-based MIAs. With those possible metrics, the adversary could decide the member and non-member by comparing metrics with thresholds. Furthermore, the combinations between the prediction probability vector and metrics are possible features for training classifiers for classifier-based MIAs, which use the classifier to distinguish members and non-members.

Regarding model types of the classifier (the attack model), we select four model types: SVM, Linear Model, XgBoost, and shallow MLP. Those models include the classical model, ensemble model, and neural network that allow detailed comparison and evaluation. Due to the utilization of varying types of classifiers, features, metrics, and relabelling operations, we implemented 54 different MIAs in total. Those MIAs are slightly different from previous MIAs and have variants to compare MIAs under the same environment and extend possible related MIAs to evaluate our newly defined metrics. The differences come from the training and settings of target models for a fair comparison, the attack model types, and the dataset usage. The extension focuses on the attack model's input features, combining the prediction probability vector with metrics for attacking with thresholds.

\subsection{Platform Implementation Details}
\label{platform_implementation_detail}

\begin{figure*}[ht!]
  \centering
  \includegraphics[scale=0.48]{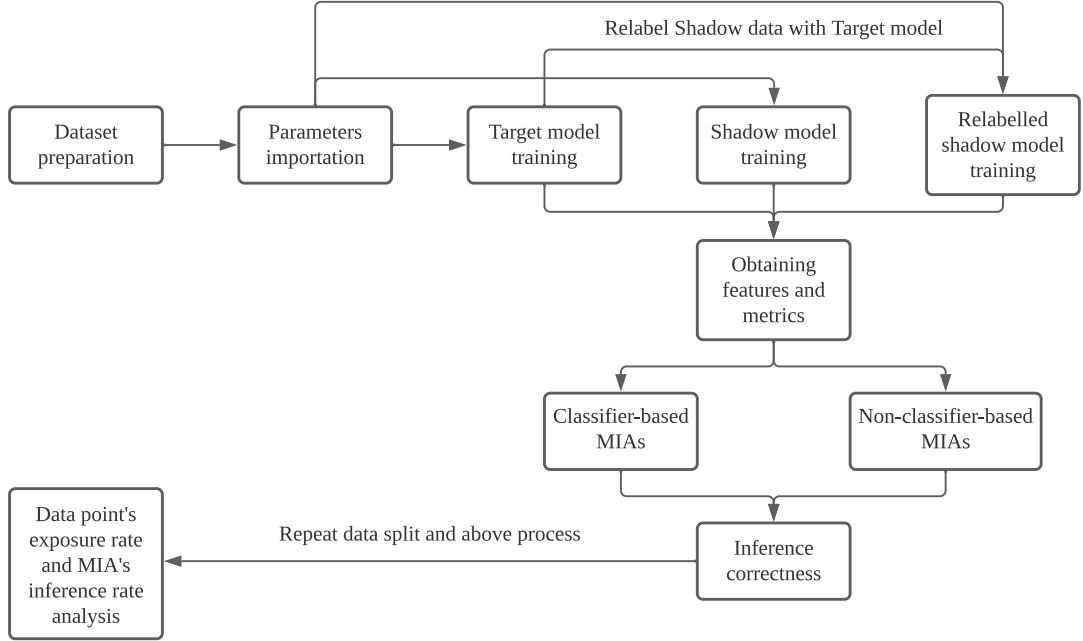}
  \caption{General process of the VMIAP.}
  \label{process_2}
\end{figure*}

We develop the VMIAP in Python and mainly use Pytorch\footnote{\url{https://pytorch.org/}} package to construct and train models. The code implementation is divided into multiple parts to make the platform scalable and flexible. Those parts comprise dataset preparation, main process treatment, parameters setting, model training, features and metrics extraction, classifier-based MIAs implementation, non-classifier-based MIAs implementation, tool package utilization, and analysis. Figure~\ref{process_2} shows the general process of the VMIAP. Dataset preparation and parameter import are necessary for training target, shadow, and relabelled shadow models. These two parts divide the dataset and prepare hyperparameters for training models, which consist of continuously updating models' weights. With the help of models, we obtain features and metrics for MIAs. The platform will record the inference correctness of multiple MIAs to each data point. The above steps are for one dataset split. After splitting the dataset several times and repeating those steps, we can analyze the data points' exposure rate and MIA's inference rate, which are related to data points' vulnerability under multiple MIAs and target models. 

The code implementation\footnote{\url{https://github.com/csong27/membership-inference}} of Shokri et al. leverages the prediction probability vector as the feature to train a binary classifier for the MIA. Salem et al. include MIAs based on prediction probability vector, transfer attack (the shadow data is not from the same distribution as the target data), and maximum probability in their Github code\footnote{\url{https://github.com/AhmedSalem2/ML-Leaks}}. Another GitHub repository\footnote{\url{https://github.com/privacytrustlab/ml\_privacy\_meter}} of Shokri et al. uses the prediction probability vector, loss, gradient, and intermediate output as features to meter the model's privacy. The MIAs in previous code implementations are fewer than our platform.
The main differentiating point is that we analyze data points' vulnerability under multiple MIAs and target models.

\section{Experimental Settings}

We designed and run multiple experiments related to three datasets. Our experiments are composed of two parts. The first part is implementing various membership inference attacks on different target models to observe data points' vulnerability under multiple MIAs and the inference tendency of the MIA to various data points. The second part compares vulnerable data points found by previous methods with vulnerable data points under multiple MIAs proposed in our paper to find overlapping vulnerable data points. In Section~\ref{datasets}, we describe datasets used for experiments. In Section~\ref{experiment_steps}, we provide the details of the experimental steps.

\subsection{Datasets}
\label{datasets}
We use three datasets: CIFAR-10, MNIST, and PURCHASE-100. We do not use all the data points in the dataset and pick fixed $40\,000$ data points from each dataset.
There are two reasons for selecting stationary $40\,000$ data points. The first reason is that fixed data points are vital for analyzing data points' vulnerability after the repetitive division of those data points for training models and attacking. The second reason is that Shokri et al. explored the performance of attacking while training target models with $10\,000$ data points. We follow their setting and choose $40\,000$ data points for the training and testing target and shadow models. Among $40\,000$ data points, non-overlapping $10\,000$ data points are used for training the target model, testing the target model, training the shadow model, and testing the shadow model, respectively. Among the datasets, the CIFAR-10 and the MNIST are almost balanced. The PURCHASE-100 is unbalanced, which means the number of data points in each class varies significantly.
Due to the CIFAR-10 and MNIST being typical datasets in image classification tasks, we only introduce PURCHASE-100. The PURCHASE-100 is based on Kaggle's "acquire valued shoppers" challenge dataset that contains shopping histories for several thousand individuals\footnote{\url{https://www.kaggle.com/c/acquire-valued-shoppers-challenge/data}}. In the work of Shokri et al., the authors use a simplified purchase dataset with 600 binary features, which represents whether a user purchases a specific item or not~\cite{shokri_membership_2017}. The PURCHASE-100 has 100 categories, which means 100 purchase types among many users. We use the same PURCHASE-100 dataset as Shokri et al.

\subsection{Experimental Steps Detail}
\label{experiment_steps}

In this section, we explain the detail of the experiments and provide the setting and architectures of models for experimental reproduction. In Section~\ref{various MIAs}, we present experimental steps, models' structures, and settings. Then, we discuss previous methods of finding vulnerable data points and how we compare vulnerable data points in Section~\ref{find_and_compare_vulnerable_data_points}.

\begin{figure*}[ht!]
  \centering
  \includegraphics[scale=0.48]{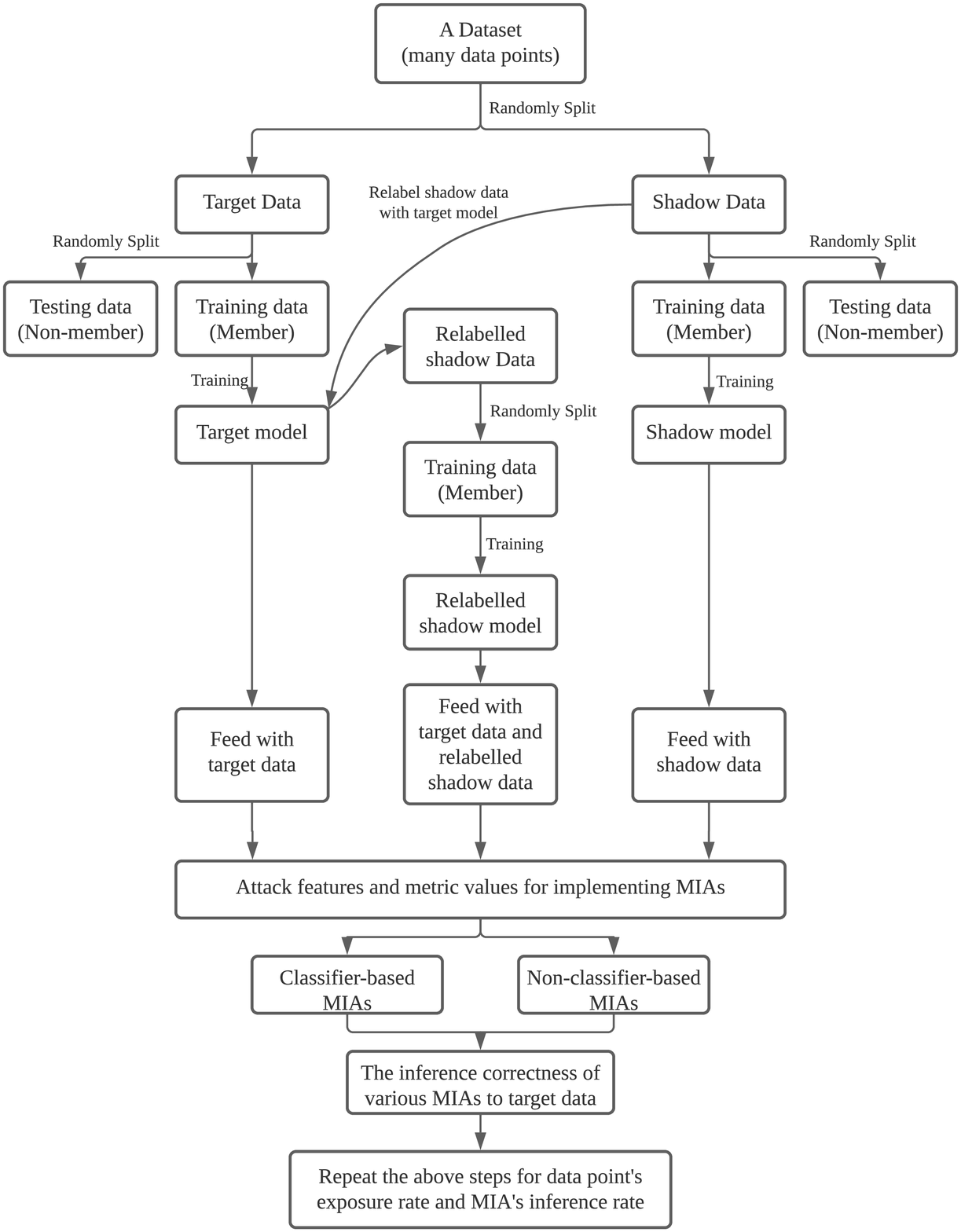}
  \caption{The process of applying various membership inference attacks}
  \label{process_1}
\end{figure*}

\subsubsection{Various Membership Inference Attacks on Different Target Models}
\label{various MIAs}

The process for each dataset is similar when applying various MIAs to different target models. First, we select $40\,000$ data points in each dataset as the total number of data points. Figure~\ref{process_1} shows each dataset's process of applying various membership inference attacks. As the figure shows, $40\,000$ data points are randomly divided into target and shadow data. The \textit{target shadow rate} means the proportion of target data. In most of the experiments, the \textit{target shadow rate} is 0.5 to guarantee the imitation effect of the shadow model. Still, this rate is different while exploring its influence on results. Then, training a target model with some randomly selected points from the target data, which are members of the target model, and the remaining data points are non-members used for testing. The \textit{target split rate} means the proportion of training data while training the target model. In most experiments, the \textit{target split rate} is 0.5 to ensure the same training and testing data for a fair evaluation. This rate is different while exploring its effect. The split of shadow data is similar to target data. The \textit{shadow split rate} gives the proportion of training data while training the shadow model. In most experiments, this rate is 0.5 for the same purpose as the \textit{target split rate}. The training setting and process of the target and shadow model are the same in our experiments. After training the target and shadow models, we relabel shadow training and testing data with the target model and obtain the relabelled shadow data used for training and testing the relabelled shadow model. The target, shadow, and relabelled shadow models are obtained during the above process.

The membership inference attack takes advantage of the difference between training and testing data points to distinguish them. To achieve this goal, we feed target data to the target model, target data, and relabelled shadow data to relabelled shadow model, and shadow data to the shadow model to obtain attack features and metric values of implementing MIAs. As mentioned before, attack features and metrics are composed of the prediction probability vector, the maximum probability value, the probability of ground truth, the cross-entropy loss, the CELoss, the Mentr value, and the entropy of the prediction vector and its normalization value. Those features and metric values are proposed by previous research work. Here, we implement them with a slight change in the VMIAP framework and add some variants.

We implement different classifier-based and non-classifier-based MIAs with attack features and metrics. After that, the inference correctness of MIAs on each data point is explicit, i.e., whether the data point is correctly inferred as a member or non-member of the target data. The above process considers applying various MIAs to a specific target model. We repeat the above process to understand the data points' situation on different target models and MIAs' inference tendency to some data points. The \textit{Split\_num} means the repetition times, representing the number of target models with the same structure, training process, and hyperparameters on a dataset. In most of the experiments, the repetition time is 20. According to the related knowledge of probability theory~\cite{Jaynes_probability_2011}, most data points will occur in training and testing data of the target model five times, which is enough to evaluate data points' vulnerability. Nevertheless, this number is different when exploring its effort on results. After that, we can analyze the data point's vulnerability under multiple MIAs and the inference tendency of the MIA to some data points.

The architecture, training process, and hyperparameters are the same for a dataset to get different target models. We choose two model architectures for one dataset to explore the influence of model architecture. The model structures' detail and training settings of each structure are explained as follows. For the MNIST dataset, components of one model structure are Conv2D (1, 4, 2), Relu, Conv2D (4, 16, 2), Relu, MaxPool2D (2), Linear ($2\,704$, $1\,024$), Dropout (0.5), and Linear ($1\,024$, 10) from input to output. Training settings are 100 for epoch number, 16 for batch size, 0.001 for learning rate, 0.01 for weight decay, Adam for optimization algorithm, and CrossEntropyLoss for loss. 
The other model structure is ResNet18, with adaptive input for the MNIST dataset. Training settings remain the same besides the weight decay change from 0.01 to 1e-7 to fit two model structures. For the PURCHASE-100 dataset, Linear (600, 128), Tanh, Linear (128, 100), Tanh, and SoftMax constitute the first model structure. The settings are 200, 16, 0.001, 1e-7, Adam, and  CrossEntropyLoss for epoch number, batch size, learning rate, weight decay, optimization algorithm, and loss. The other model structure contains Linear (600, 512), Tanh, Dropout (0.3), Linear (512, 256), Tanh, Dropout (0.3), Linear (256, 128), Tanh, Dropout (0.3), Linear (128, 100), Tanh, and SoftMax from input to output. Training settings are the same as the first model structure of PURCHASE-100. 
For the CIFAR-10 dataset, LeNet is the first model structure. Training settings are 100 for epoch number, 36 for batch size, 0.001 for learning rate, 1e-7 for weight decay, Adam for optimization algorithm, and CrossEntropyLoss for loss. The second model structure is ResNet18 and is trained with the same settings as the first model. Those model architectures are commonly used for image classification tasks on selected datasets~\cite{Lecun_letnet_1998,He_resnet_2016}. We attempt to evaluate various model architectures in our experiments, and we determine the architectures of models based on previous works~\cite{shokri_membership_2017,salem_ml-leaks_2018} and evaluation factors, like overfitting level.

\subsubsection{Finding and Comparing Vulnerable Data Points}
\label{find_and_compare_vulnerable_data_points}

As mentioned in Section~\ref{exposure rate}, we can determine vulnerable data points with the help of AMER and ANMER. We define those two metrics according to the actual inference correctness of MIAs against target models. To compare and test the effectiveness of previous methods under multiple MIAs and target models, we pick 40 data points out of $40\,000$ total data points as vulnerable data points based on newly defined metrics and previous evaluation methods. Partial reasons for selecting 40 data points are that the number of vulnerable data points is relatively small, and the fixed number of vulnerable data points in each approach is easier to compare. Besides, if the correctly inferred data point sets differ among 54 MIAs against 20 (\textit{Split\_num}) target models trained with different data splits, the value 40 (slightly larger than $40\,000$/(20*54)$\approx$37) can observe this worst case and is rounding for convenient calculation. The steps of finding vulnerable data points are identical among various datasets, and the following paragraphs describe how we use different methods to determine vulnerable data points.

\textbf{Our new metrics.} Following the definitions, we compute the AMER of each data point and select 40 vulnerable data points based on this value in each model structure of the dataset. Then, we change the evaluation metric to the ANMER and can also obtain 40 vulnerable data points. Besides, we repeat training \textit{Split\_num} target models, attacking with 54 MIAs and obtaining vulnerable data points with our new metrics. The reason is that we try to observe the difference between vulnerable data points while repeating the experiment. Thus, we can get four vulnerable data point sets, each of which has 40 vulnerable data points for each dataset. 

\textbf{Neighbors-based method.} Long et al.~\cite{long_pragmatic_2020} selected data points with fewer neighbors on the combination feature space represented by outputs of some reference models as vulnerable ones. We use $40\,000$ data points in the empirical experiments as the data for target models (target data) and the remaining data points for reference models (reference data). Then, we construct ten reference datasets, each of whose size is $10\,000$ from reference data with replacement to train ten reference models. We can obtain the combined outputs of ten reference models to one data point in target data and part of reference data as its combination feature. After that, the distance between each target point and reference data is computed via the cosine distance of their combination features. Setting the neighbor threshold between neighbors as 0.1 (a setting in the original paper) and increasing the probability threshold from the minimum to the maximum value, we can obtain 40 more vulnerable data points from the target data. The final point is that we repeat the above process without changing anything and get two vulnerable data point sets with this method.

\textbf{SUOD.} The SUOD~\cite{zhao_suod_2020}, which expedites the training and prediction with many unsupervised detection models, is a three-model acceleration framework to detect abnormal data points. The SUOD comprises of a Random Projection module, a Balanced Parallel Scheduling module, and a Pseudo-supervised Approximation module. The main task of the Random Projection module is to reduce feature dimensions and keep distance relationships. A balanced Parallel Scheduling module can forecast models' training and prediction costs with high confidence. The data points found with the SUOD might not be the real vulnerable ones. We use the data points found by this method to compare with other methods. The last step is repeating the above process without changing anything to get two vulnerable data point sets with this method.

\textbf{Privacy risk score.} Song et al.~\cite{song_systematic_2021} defined the privacy risk score of a single data point as its probability of being a member under the condition that the target model's output on this point is observed when the target model is attacked by their new-defined MIA. The new-defined MIA is a non-classifier-based MIA that uses the Mentr value they proposed with different value thresholds for each category. For each target model, even the similar target models, which are only different at dataset split, the Mentr values are different. We find vulnerable data points whose privacy risk scores are higher for each target model with Song's code implementation\footnote{\url{https://github.com/inspire-group/membership-inference-evaluation}}. Due to multiple target models being trained, the corresponding number of vulnerable data point sets are obtained. We count the appearance of each data point and select 40 more frequent data points as vulnerable data points found with this method. Again, we repeat the above process without changing anything. 

\textbf{Shapley value.} In the work of Duddu et al.~\cite{duddu_shapr_2021}, the authors proposed to measure data points' susceptibility to MIAs with the Shapley value of each data point. The Shapley value represents a data point's influence on the target model's output predictions. Unlike the original way of computing data points' Shapley values by training many models to measure their impact on testing data's accuracy, they followed the approach put forward by Jia et al.~\cite{jia_scalability_2021}. This method uses a K-NN classifier to compute the contribution of training data points to a single testing data point and aggregate the single training data point's contribution to all testing data points as its Shapley value. We feed the outputs of the target model to training and testing data into a K-NN classifier and obtain training data points' Shapley values with the help of Duddu's code implementation\footnote{\url{https://github.com/AI-secure/Shapley-Study}}. Due to multiple target models being trained, a corresponding number of vulnerable data point sets with a higher Shapley value are obtained. We count the appearance of each data point and select 40 more frequent data points as vulnerable data points found with this method. Again, we repeat the above process without changing anything.

\section{Results and Discussion}

In this section, we present our experimental results and discuss the findings. In particular, we include parts about the accuracy of target models, attack performance of MIAs (Section~\ref{accuracy and performance}), exposure rate of data point (Section~\ref{exposure rate}), inference rate of MIA (Section~\ref{inference rate}), and vulnerable data points comparison (Section~\ref{vulnerable data points}).

\subsection{Target Models' Accuracy and MIAs' Attack Performance}
\label{accuracy and performance}

Table~\ref{target_and_attack_performance} shows some statistical metrics about target models' setting, accuracy, and MIAs' attack performance. Data in each row represents the result for a given configuration (\textit{Dataset}, \textit{Structure}, \textit{Split\_num}, and \textit{Rates}). We train target models (the quantity is \textit{Split\_num}) with varying data splits of the same dataset. They have identical training processes and hyperparameters under one configuration (dataset, model structure, number of splits, and rates). The \textit{Rates} column consists of three elements, i.e., \textit{target shadow rate}, \textit{target split rate}, and \textit{shadow split rate} in Section~\ref{experiment_steps}, to divide the dataset into four sub-datasets for training and testing target and shadow models. Besides, we calculate the average training accuracy (\textit{Avg train acc}), the average testing accuracy (\textit{Avg test acc}), and the average accuracy difference between training and testing accuracy (\textit{Avg dif acc}) of target models.

\begin{table*}[t!]
\caption{Target models' settings, accuracy, and MIAs' attack performance.}
\label{target_and_attack_performance}
\begin{adjustbox}{max width=1\textwidth}
\begin{tabular}{cc|c|c|c|c|c|c|c|c|c|c|c|c|}
\cline{2-14}
 & \multicolumn{4}{|c|}{Configurations}& \multicolumn{9}{c|}{Metrics}\\
\hline

\multicolumn{1}{|c|}{Row} & Dataset & Structure  & Split\_num & \begin{tabular}[c]{@{}c@{}}Rates \\(target shadow rate, \\target split rate,
\\shadow split rate)\end{tabular}& \begin{tabular}[c]{@{}c@{}}Avg train\\ acc\end{tabular} & \begin{tabular}[c]{@{}c@{}}Avg test\\ acc\end{tabular} & \begin{tabular}[c]{@{}c@{}}Avg dif\\ acc\end{tabular} & \begin{tabular}[c]{@{}c@{}}Avg MIA\\ acc\end{tabular} & \begin{tabular}[c]{@{}c@{}}MIA acc\\ var\end{tabular} & \begin{tabular}[c]{@{}c@{}}MIA acc\\ med\end{tabular} & \begin{tabular}[c]{@{}c@{}}MIA acc\\ max\end{tabular} & \begin{tabular}[c]{@{}c@{}}MIA acc\\ min\end{tabular} & \begin{tabular}[c]{@{}c@{}}Difference\\ MIA acc \\ (max - min)\end{tabular} \\
\hline
\multicolumn{1}{|c|}{1} & CIFAR-10    & LeNet      & 20         & (0.5, 0.5, 0.5) & 0.680         & 0.449        & 0.2310      & 0.532       & 0.00157     & 0.515       & 0.621       & 0.500    & 0.121   \\
\hline
\multicolumn{1}{|c|}{2} & CIFAR-10    & LeNet      & 40         & (0.5, 0.5, 0.5) & 0.680         & 0.447        & 0.2330      & 0.533       & 0.00157     & 0.515       & 0.622       & 0.500    & 0.122     \\
\hline
\multicolumn{1}{|c|}{3} & CIFAR-10    & LeNet      & 40         & (0.5, 0.8, 0.8) & 0.6504                                                  & 0.476                                                  & 0.1740                                                & 0.709                                                 & 0.01369                                               & 0.779                                                 & 0.810                                                 & 0.459  & 0.351                                               \\
\hline
\multicolumn{1}{|c|}{4} & CIFAR-10    & LeNet      & 40         & (0.8, 0.5, 0.5) & 0.6549                                                  & 0.477                                                  & 0.1778                                                & 0.527                                                 & 0.00098                                               & 0.510                                                 & 0.592                                                 & 0.500      &0.092                                           \\
\hline
\multicolumn{1}{|c|}{5} & CIFAR-10    & ResNet18   & 20         & (0.5, 0.5, 0.5) & 0.821         & 0.624        & 0.1974      & 0.621       & 0.00505     & 0.657       & 0.719       & 0.499     & 0.22  \\
\hline
\\

\hline
\multicolumn{1}{|c|}{6} & MNIST    & CNN        & 20         & (0.5, 0.5, 0.5) & 0.969         & 0.944        & 0.0255      & 0.502       & 0.000009    & 0.501       & 0.512       & 0.499   & 0.013    \\
\hline
\multicolumn{1}{|c|}{7} & MNIST    & CNN        & 40         & (0.5, 0.5, 0.5) & 0.889         & 0.874        & 0.0146      & 0.501       & 0.000002    & 0.500       & 0.504       & 0.499   & 0.005    \\
\hline
\multicolumn{1}{|c|}{8} & MNIST    & CNN        & 40         & (0.5, 0.8, 0.8) & 0.882                                                   & 0.868                                                  & 0.0142                                                & 0.7456                                                & 0.010666                                              & 0.799                                                 & 0.800                                                 & 0.499 & 0.301                                                \\
\hline
\multicolumn{1}{|c|}{9} & MNIST    & CNN        & 40         & (0.8, 0.5, 0.5) & 0.893                                                   & 0.882                                                  & 0.0105                                                & 0.5005                                                & 0.000001                                              & 0.500                                                 & 0.503                                                 & 0.499 & 0.004                                                \\
\hline
\multicolumn{1}{|c|}{10} & MNIST    & ResNet18   & 20         & (0.5, 0.5, 0.5) & 0.995         & 0.980        & 0.0157      & 0.505       & 0.000018    & 0.504       & 0.515       & 0.499  & 0.016     \\
\hline
     \\
\hline
\multicolumn{1}{|c|}{11} & PURCHASE-100 & Shallow MLP & 20         & (0.5, 0.5, 0.5) & 0.951         & 0.591        & 0.360       & 0.699       & 0.01846     & 0.739       & 0.862       & 0.498   & 0.364   \\
\hline
\multicolumn{1}{|c|}{12} & PURCHASE-100 & Shallow MLP & 40         & (0.5, 0.5, 0.5) & 0.949         & 0.588        & 0.362       & 0.700       & 0.01838     & 0.735       & 0.864       & 0.499  & 0.365      \\
\hline
\multicolumn{1}{|c|}{13} & PURCHASE-100& Shallow MLP & 40         & (0.5, 0.8, 0.8) & 0.957                                                   & 0.648                                                  & 0.309                                                 & 0.792                                                 & 0.02023                                               & 0.833                                                 & 0.868                                                 & 0.199    & 0.669                                             \\
\hline
\multicolumn{1}{|c|}{14} & PURCHASE-100 & Shallow MLP & 40         & (0.8, 0.5, 0.5) & 0.955                                                   & 0.646                                                  & 0.309                                                 & 0.637                                                 & 0.00902                                               & 0.678                                                 & 0.751                                                 & 0.499   & 0.252                                              \\
\hline
\multicolumn{1}{|c|}{15} & PURCHASE-100 & Deeper MLP & 20         & (0.5, 0.5, 0.5) & 0.787         & 0.610        & 0.177       & 0.542       & 0.00126     & 0.545       & 0.598       & 0.499  & 0.099\\
\hline
\end{tabular}
\end{adjustbox}
\end{table*}

Table~\ref{target_and_attack_performance} shows some observations about average accuracy of target models. First, the larger the gap between the training and testing accuracy of the target model, the better MIAs' attack performance, generally. The accuracy gap of MNIST (0.01$\sim$0.02) is smaller than CIFAR-10 (0.17$\sim$0.23) and PURCHASE-100 (0.17$\sim$0.36). However, the attack performance of CIFAR-10 and PURCHASE-100 is relatively higher than MNIST. We attribute it to the fact that a larger accuracy gap means a higher level of overfitting, which is the commonly accepted reason for MIA. Second, only increasing the \textit{Split\_num} or altering \textit{Rates} does not bring too much change to average training and testing accuracy. For example, in the table, the first and second rows about the results of CIFAR-10 (the difference is \textit{Split\_num}), the eleventh and twelfth rows about the results of PURCHASE-100 (the difference is \textit{Split\_num}), etc. The reason for this observation is the stability of the optimization algorithm for training machine learning models. Changing the number of splits and rates has little effort on the prediction performance of target models. However, the average attack accuracy increases while applying the \textit{Rates} of (0.5, 0.8, 0.8). Because the percentage of training data points increases, the attack model can obtain high accuracy if it infers most data points as members while training the attack model and using it for attacking. It means that the changing of \textit{Rates} does not affect the attack ability of MIA and is an external disturbance. Additionally, the \textit{Split\_num} has little effort on the attack performance.



\begin{figure*}
     \centering
     \begin{subfigure}[b]{0.5\textwidth}
         \centering
         \includegraphics[width=\textwidth]{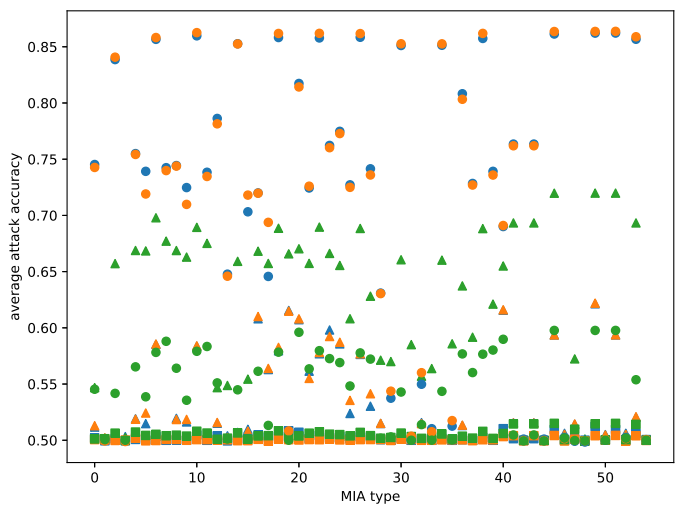}
         \caption{Average Attack Accuracy}
         \label{performance}
     \end{subfigure}
     \begin{subfigure}[b]{0.45\textwidth}
         \centering
         \raisebox{0.8in}{\includegraphics[width=\textwidth]{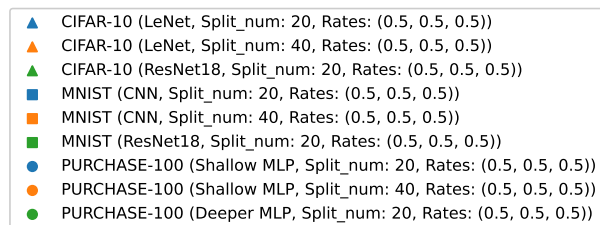}}
         \caption{Legend}
     \end{subfigure}
        \caption{The average attack accuracy of different MIAs. In Figure~\ref{performance}, the x-axis represents different MIAs. The y-axis indicates the average attack accuracy over multiple splits. Various colors and shapes show varying configurations.}
        \label{attack_performance}
\end{figure*}

Regarding the attack accuracy, we compute the average attack accuracy over different splits (or target models) to represent a specific MIA's attack performance for those target models. As Figure~\ref{attack_performance} shows, each point in the picture means the average attack accuracy of a configuration (dataset, model structure, number of splits, and rates) under one MIA. Due to 54 different MIAs, we obtain a list of attack performance values, which are points with the same color and shape. Then, we calculate the mean (\textit{Avg MIA acc}), variance (\textit{MIA acc var}), median (\textit{MIA acc med}), maximum (\textit{MIA acc max}), and minimum (\textit{MIA acc min}) values of each list of attack performance values in Table~\ref{target_and_attack_performance}.

In Figure~\ref{attack_performance} and Table~\ref{target_and_attack_performance}, we can see that the average attack accuracy of MIAs ranges from 0.5 to 0.9 among the three datasets. However, the average attack accuracy variance is relatively small (0.000002$\sim$0.02) within the same configuration (dataset, model structure, number of splits, and rates). This phenomenon shows the configuration will significantly affect the attack performance. According to the previous analysis, the \textit{Rates} does not affect the attack ability. The \textit{Split\_num} has little effort on the attack performance. Therefore, the dataset and model are non-negligible factors of MIA's attack performance apart from the accuracy gap. This conclusion corresponds with the result of previous work~\cite{truex_towards_2018}. From the last column of the table, we can observe a significant difference between the maximal and minimal average attack accuracy of 54 MIAs. For example, row 12 about the result of PURCHASE-100 has a difference of 0.365. At the same time, the minimal average attack accuracy (\textit{MIA acc min}) is close to 0.5 in all experiments. This result indicates that not all MIAs are efficient at attacking target models under varying configurations. The MIA, which achieves a higher attack performance than other MIAs under one configuration, may not obtain competitive attack performance in other configurations compared with other MIAs. It gives us the inspiration that we need to select suitable MIA for a specific combination of dataset and model structure to get high attack performance.



\subsection{Data Point's Exposure Rate}
\label{exposure rate}

Following Section~\ref{definitions}, the data point's AMER and ANMER represent its vulnerability under multiple MIAs and target models. The difference between those two metrics is the data point is in the training or testing data of target models. As the explanation in Section~\ref{find_and_compare_vulnerable_data_points} for finding vulnerable data points, we leverage those two metrics for selecting vulnerable data points under multiple MIAs and target models in empirical results. This section focuses on analyzing those two metrics and their difference.

\begin{figure*}
     \centering
     \begin{subfigure}[b]{0.2\textwidth}
         \centering
         \includegraphics[width=\textwidth]{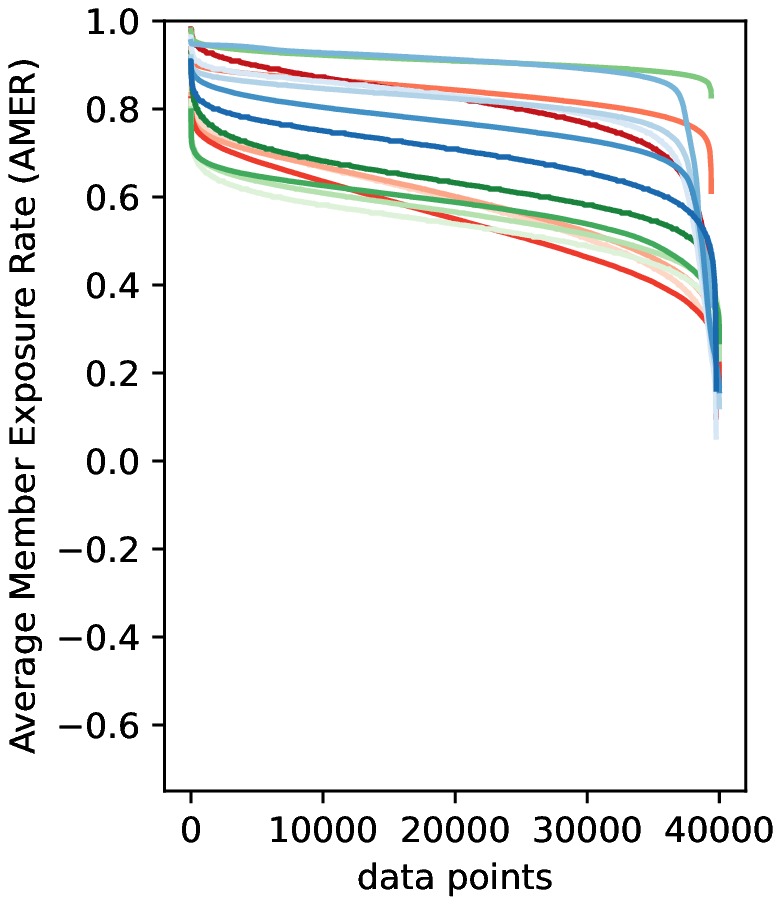}
         \caption{AMER}
     \end{subfigure}
     \begin{subfigure}[b]{0.2\textwidth}
         \centering
         \includegraphics[width=\textwidth]{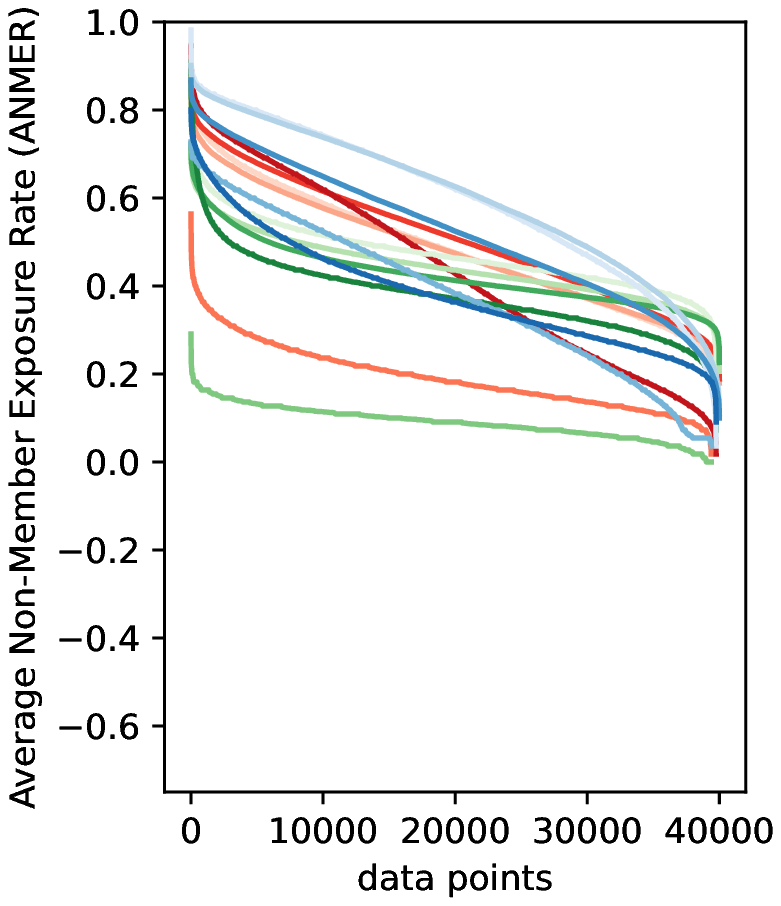}
         \caption{ANMER}
     \end{subfigure}
     \begin{subfigure}[b]{0.2\textwidth}
         \centering
         \includegraphics[width=\textwidth]{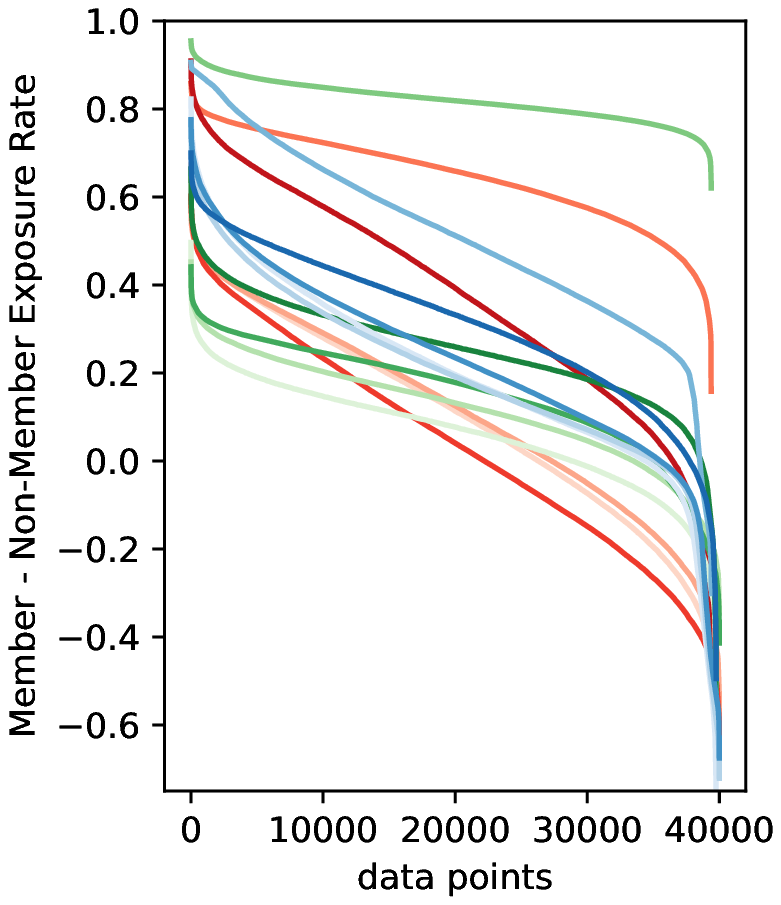}
         \caption{Difference}

     \end{subfigure}
     \begin{subfigure}[b]{0.3\textwidth} 
         \centering
         \raisebox{0.2in}{\includegraphics[width=\textwidth]{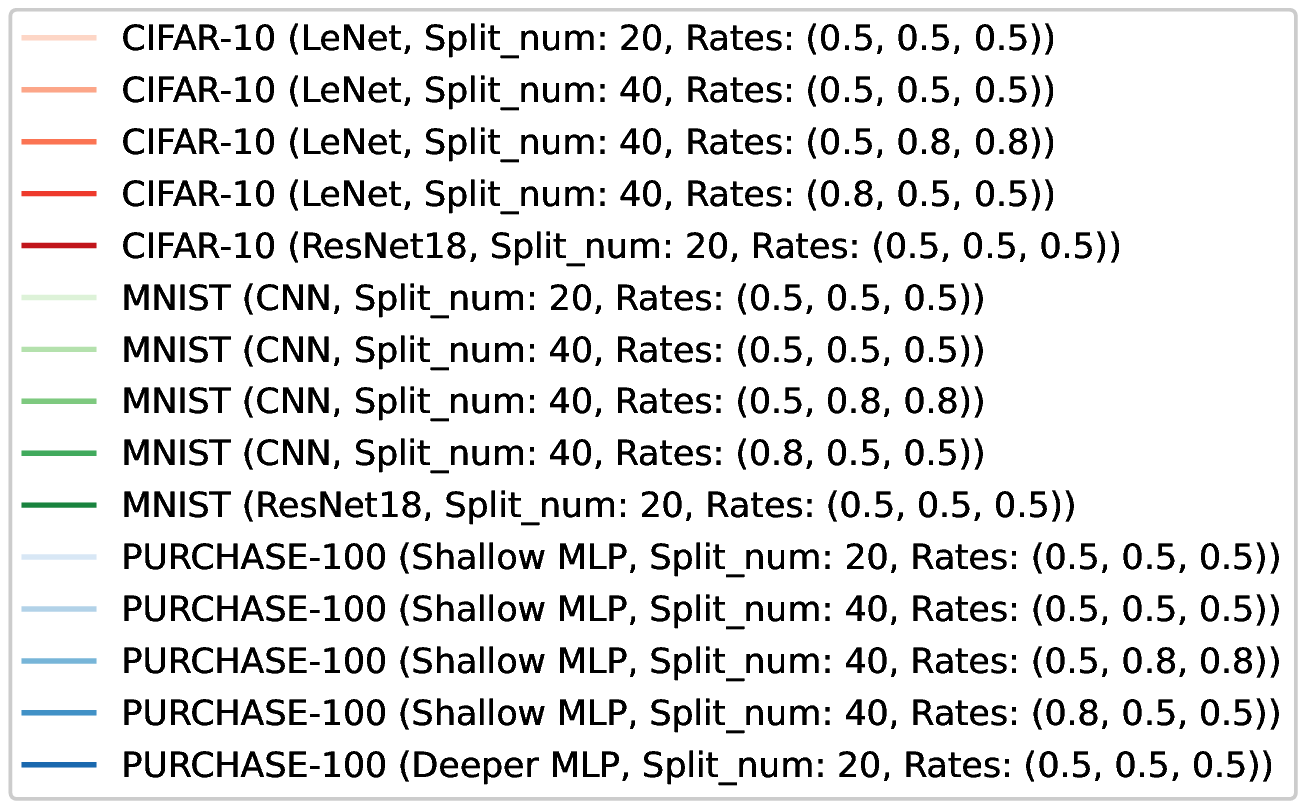}}
         \caption{Legend}

     \end{subfigure}
        \caption{The sorted AMER values, ANMER values, and difference of those two metrics under varying configurations. The definitions of AMER and ANMER are provided in equations~\eqref{AMER_formula} and~\eqref{ANMER_formula}. Different lines in each picture represent varying configurations, which are also explained in Table~\ref{target_and_attack_performance}. Each line is drawn as the decrease of the corresponding value (AMER, ANMER, and Difference). The x-axis indicates the symbol of data points. As there is no direct relation between lines, the same $x$ symbol usually means different data points in varying lines because of value sorting. }
        \label{Three_pics_about_exposure_rate}
\end{figure*}

Figure~\ref{Three_pics_about_exposure_rate} shows the sorted AMER values, ANMER values, and the difference between those two metrics under varying configurations (dataset, model structure, number of splits, and rates). There are some observations about the metric values in Figure~\ref{Three_pics_about_exposure_rate}. First, we can see that the AMER and ANMER decrease from a value close to 1 to a value near 0. The ANMER has a faster and more significant drop than the AMER (from the shape and terminal of curves), which leads to a large proportion of the AMER value being higher than the ANMER value. This observation is conceivable because MIA is biased toward inferring the training data rather than the testing data. It is why we do not call the attack a Non-Membership Inference Attack. Therefore, training data points' vulnerability is more remarkable than testing data points under multiple MIAs. It corresponds with the result that the AMER value is higher than the ANMER value. From another perspective, our newly defined metrics can reflect the actual situation of data points' vulnerability under multiple MIAs and target models. Second, there is a ratio of data points with an AMER or ANMER value close to 1, meaning data points are almost correctly inferred by 54 MIAs while being in different target models. Therefore, a high AMER or ANMER value reflects that the data point is vulnerable. Specifically, around 25 percent of data points have an AMER value larger than 0.6 in all experiments in Figure~\ref{Three_pics_about_exposure_rate}.
Meanwhile, a small part of data points has an ANMER value larger than 0.6. Those results indicate that there are indeed partial vulnerable data points under multiple MIAs and target models. Our new metrics, AMER and ANMER, can reflect and capture vulnerable data points under multiple MIAs and target models. Third, the AMER and ANMER values change as the alteration of split rates. In particular, the AMER values are high when the split rates are 0.5, 0.8, and 0.8. The AMER values are relatively high when the split rates are 0.5, 0.5, and 0.5. The possible reason for this phenomenon is that more training data points lead to the class imbalance of the dataset for training the attack model. The attack model tends to predict all data points to members. Therefore, the AMER values are higher. Fourth, only increasing the \textit{Split\_num} from 20 to 40 does not change too much to the curves of the AMER and ANMER values. It reflects that those two metrics are not sensitive to the \textit{Split\_num}. The \textit{Split\_num} represents the number of target models whose training or testing data might contains a specific data point. Increasing the \textit{Split\_num} means enlarging the number of target models whose training data includes this specific data point while calculating its AMER value. The sum of MER values also increases because the AMER value is the average value of MER among target models whose training data includes this specific data point (Section~\ref{exposure rate}). Therefore, the \textit{Split\_num} does not affect AMER and ANMER values much.


\begin{figure}
     \centering
     \begin{subfigure}[b]{0.22\textwidth}
         \centering
         \includegraphics[scale=0.45]{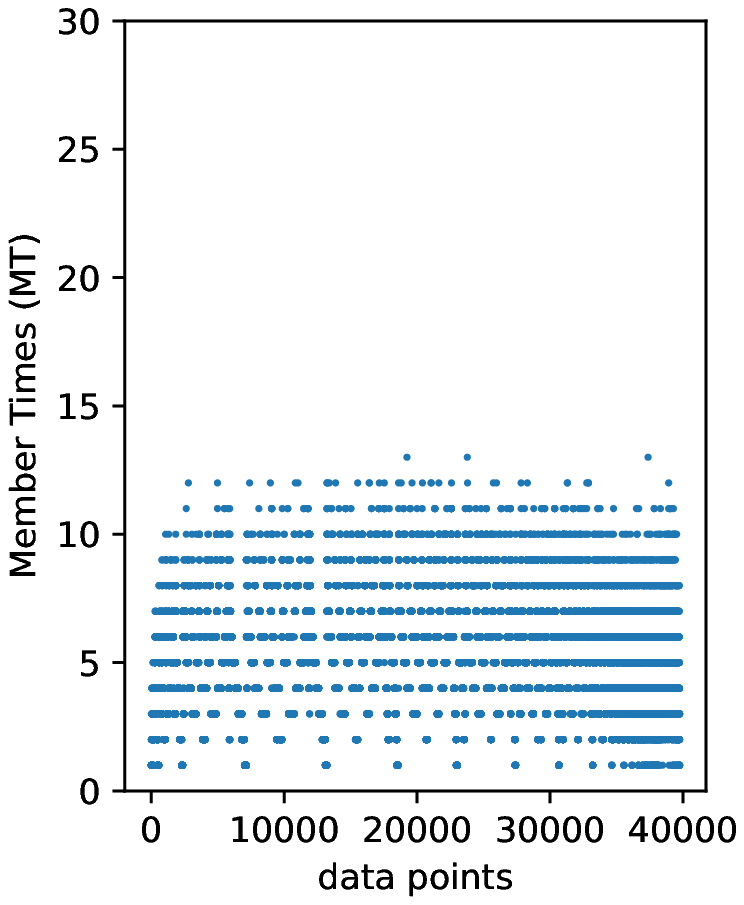}
         \caption{Less Splits}
         \label{s_20}
     \end{subfigure}
     \begin{subfigure}[b]{0.22\textwidth}
         \centering
         \includegraphics[scale=0.45]{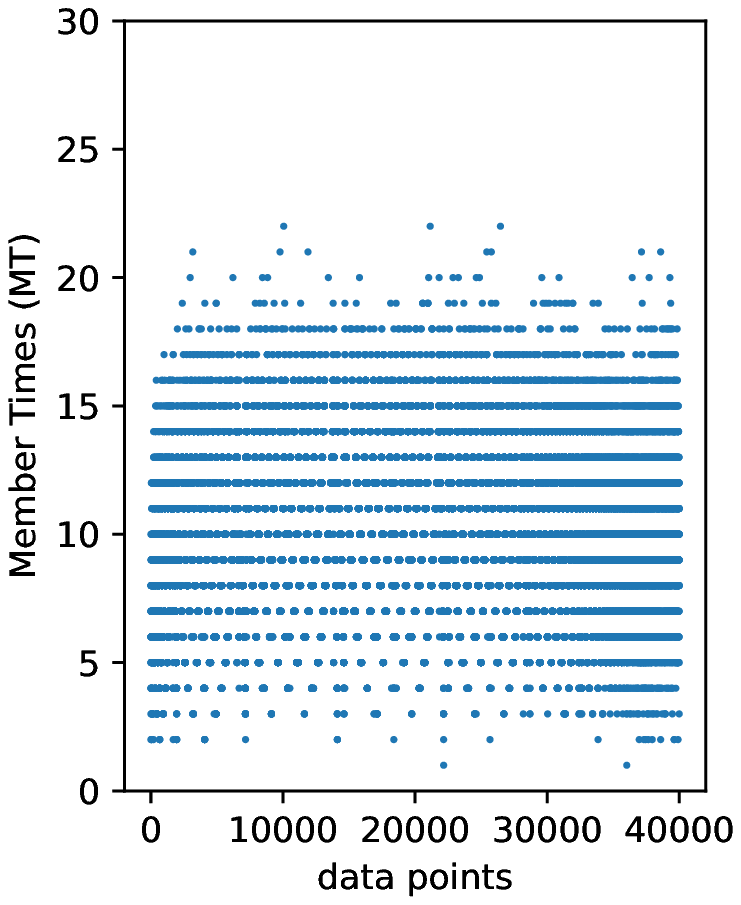}
         \caption{More Splits}
         \label{s_40}
     \end{subfigure}
        \caption{The MT of data points as the decrease of AMER values in two configurations. The only difference between Figure~\ref{s_20} and Figure~\ref{s_40} is the Split\_num (20 and 40, respectively), which is the number of splits or target models for the configuration (PURCHASE-100, LeNet, and Rates (0.5, 0.5, 0.5)).}
        \label{Member-time_for_AMER}
\end{figure}

\begin{figure}
     \centering
     \begin{subfigure}[b]{0.22\textwidth}
         \centering
         \includegraphics[scale=0.45]{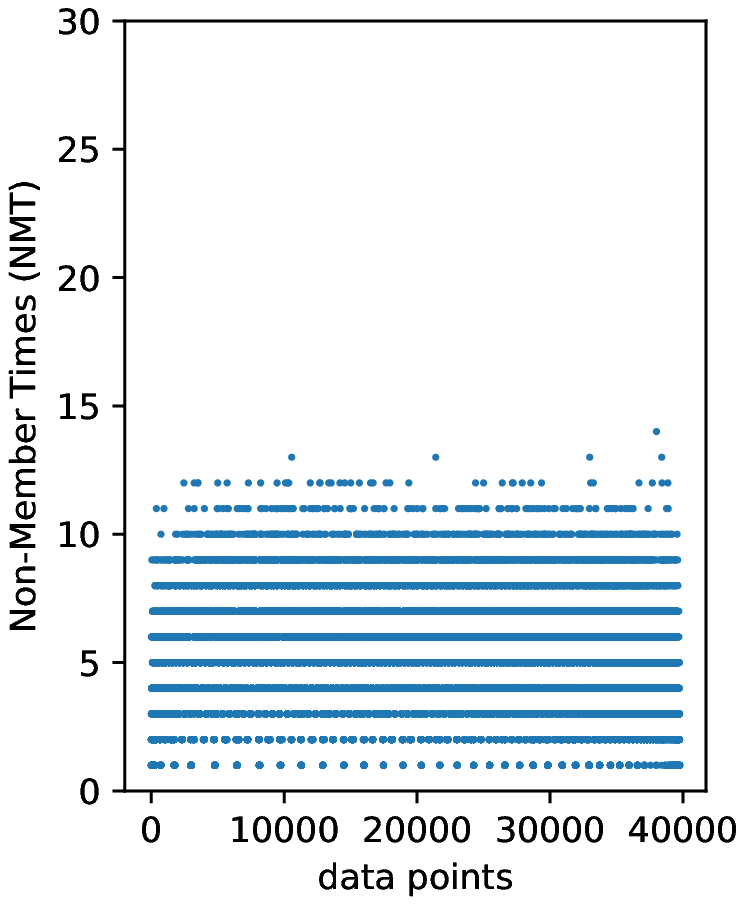}
         \caption{Less Splits}
         \label{sn_20}
     \end{subfigure}
     \begin{subfigure}[b]{0.22\textwidth}
         \centering
         \includegraphics[scale=0.45]{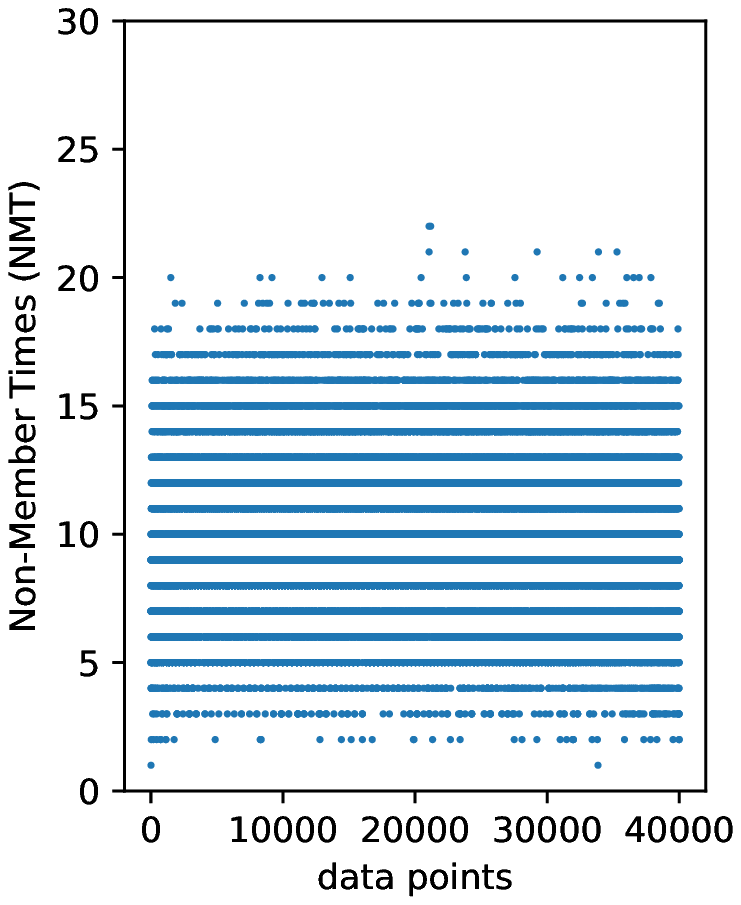}
         \caption{More Splits}
         \label{sn_40}
     \end{subfigure}
        \caption{The NMT of data points as the decrease of ANMER values in two configurations. The only difference between Figure~\ref{sn_20} and Figure~\ref{sn_40} is the Split\_num (20 and 40, respectively), which is the number of splits or target models for the configuration (PURCHASE-100, LeNet, and Rates (0.5, 0.5, 0.5)).}
        \label{Non-member-time_for_ANMER}
\end{figure}

\begin{figure}
     \centering
     \begin{subfigure}[b]{0.22\textwidth}
         \centering
         \includegraphics[scale=0.45]{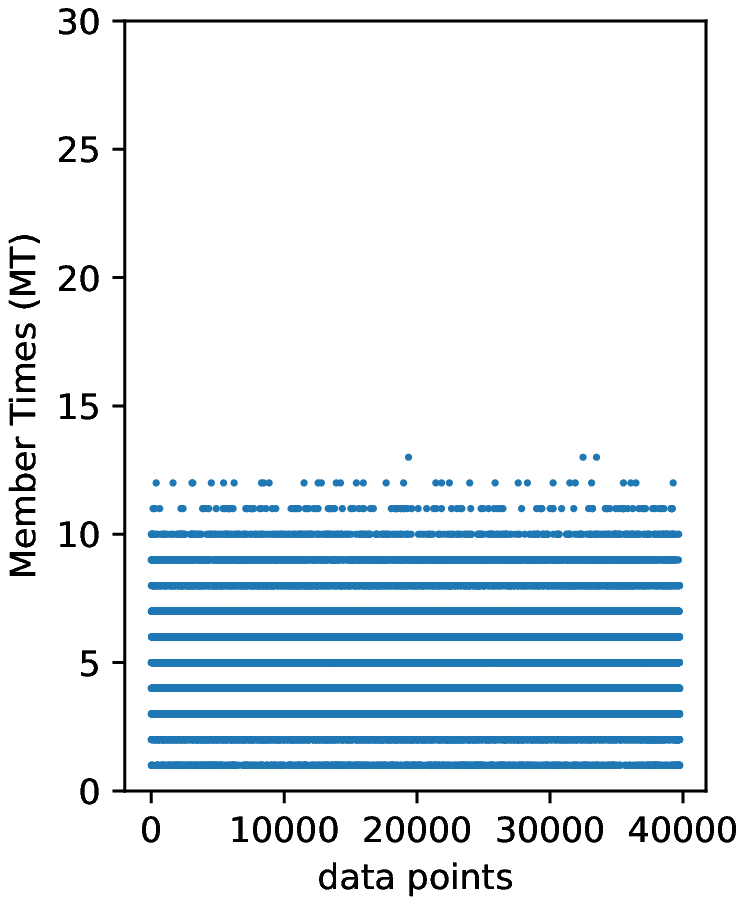}
         \caption{MT}
         \label{mt_f}
     \end{subfigure}
     \begin{subfigure}[b]{0.22\textwidth}
         \centering
         \includegraphics[scale=0.45]{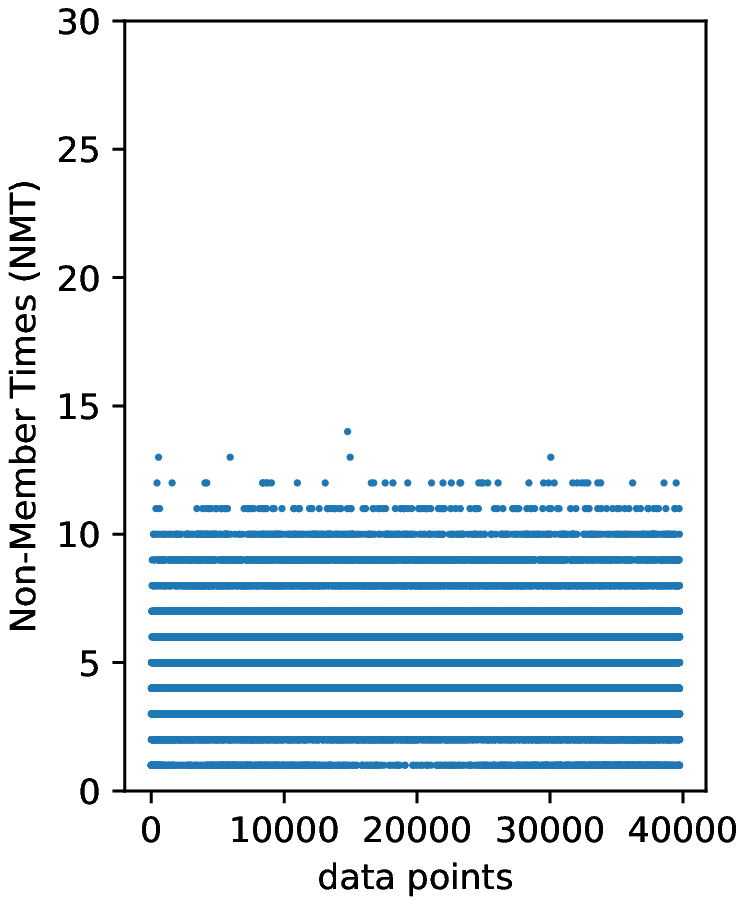}
         \caption{NMT}
         \label{nmt_f}
     \end{subfigure}
        \caption{The MT and NMT of data points as the decrease of difference between the AMER and ANMER values in one configuration. The configuration is PURCHASE-100 with the settings (Shallow MLP, Split\_num: 20, Rates: (0.5, 0.5, 0.5)).}
        \label{Member_and_Non-member_for_AMER_and_ANMER_comparison}
\end{figure}

Due to limited random splits of the dataset for experiments, the MT and NMT of different data points vary, which is inevitable because it is impossible to iterate all the splits of the dataset with many data points. Figure~\ref{Member-time_for_AMER} shows the MT of data points as the decrease of AMER values. Figure~\ref{Non-member-time_for_ANMER} shows the NMT of data points as the decrease of ANMER values. Figure~\ref{Member_and_Non-member_for_AMER_and_ANMER_comparison} shows MT and NMT of data points as the decrease of difference between the AMER and ANMER values. Each blue point in those three figures represents the MT or NMT of one data point. In Figure~\ref{Member-time_for_AMER}, there is a slightly sparse area at the bottom of each picture. Still, the overall distribution of MT over all data points is even. Similarly, the general distributions of MT or NMT in Figure~\ref{Non-member-time_for_ANMER} and Figure~\ref{Member_and_Non-member_for_AMER_and_ANMER_comparison} are balanced over all the data points. According to equations~\eqref{AMER_formula} and~\eqref{ANMER_formula}, the MT and NMT influence the AMER and ANMER metrics calculation. The distributions of MT and NMT are even in the three mentioned figures. However, AMER and ANMER values for each data point still vary greatly from curves in Figure~\ref{Three_pics_about_exposure_rate}. It means values of MT and NMT among data points are not the reason for data points' fluctuating AMER and ANMER metrics under our experiments. We can obtain similar figures and observations from the left experiments with different configurations. We do not display all of them due to limited space.



\subsection{MIA's Inference Rate}
\label{inference rate}

\begin{figure*}
     \centering
     \begin{subfigure}[b]{1.0\textwidth}
         \centering
         \includegraphics[scale=0.6]{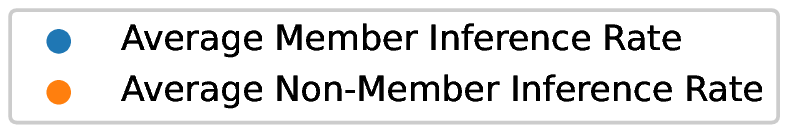}
         \caption{Legend}
     \end{subfigure}
     \begin{subfigure}[b]{0.3\textwidth}
         \centering
         \includegraphics[scale=0.6]{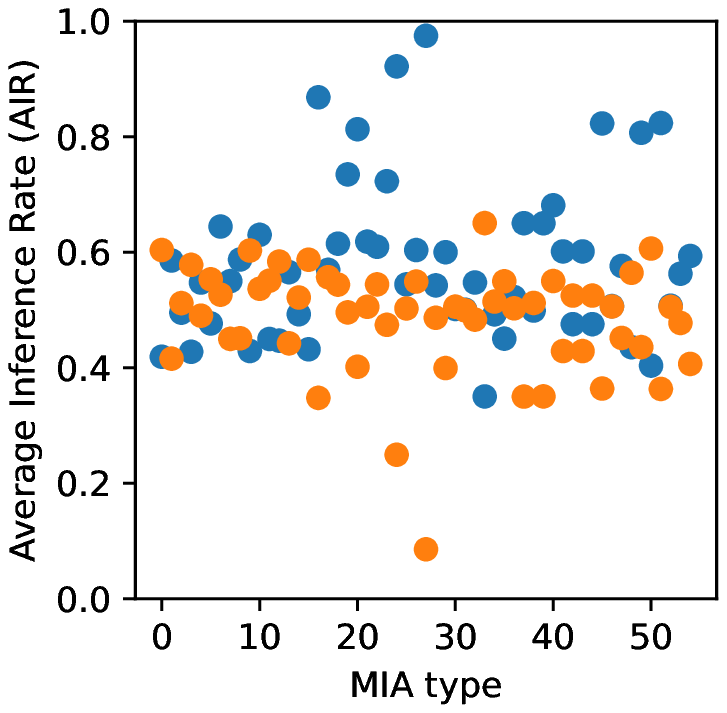}
         \caption{CIFAR-10, LeNet, Split\_num: 20, Rates: (0.5, 0.5, 0.5)}
     \end{subfigure}
     \hfill
     \begin{subfigure}[b]{0.3\textwidth}
         \centering
         \includegraphics[scale=0.6]{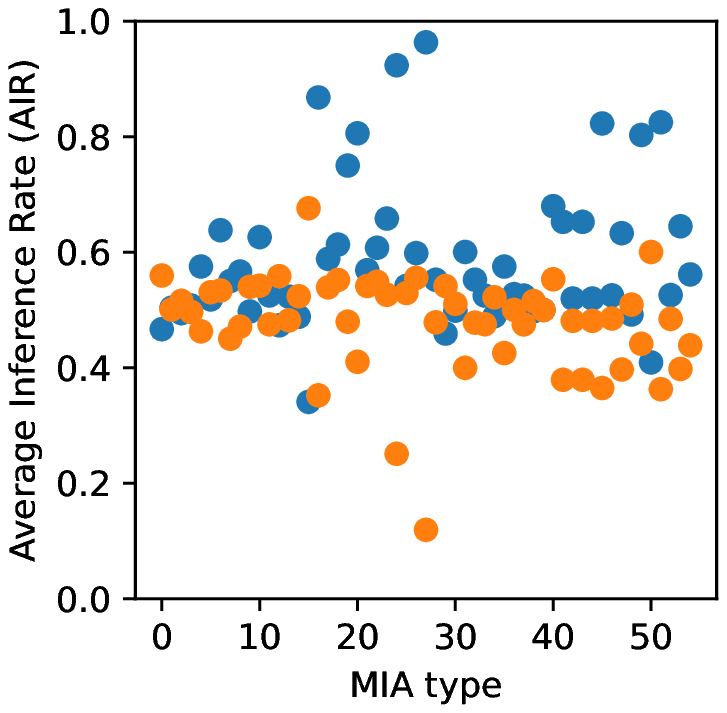}
         \caption{CIFAR-10, LeNet, Split\_num: 40, Rates: (0.5, 0.5, 0.5)}
     \end{subfigure}
     \hfill
     \begin{subfigure}[b]{0.3\textwidth}
         \centering
         \includegraphics[scale=0.6]{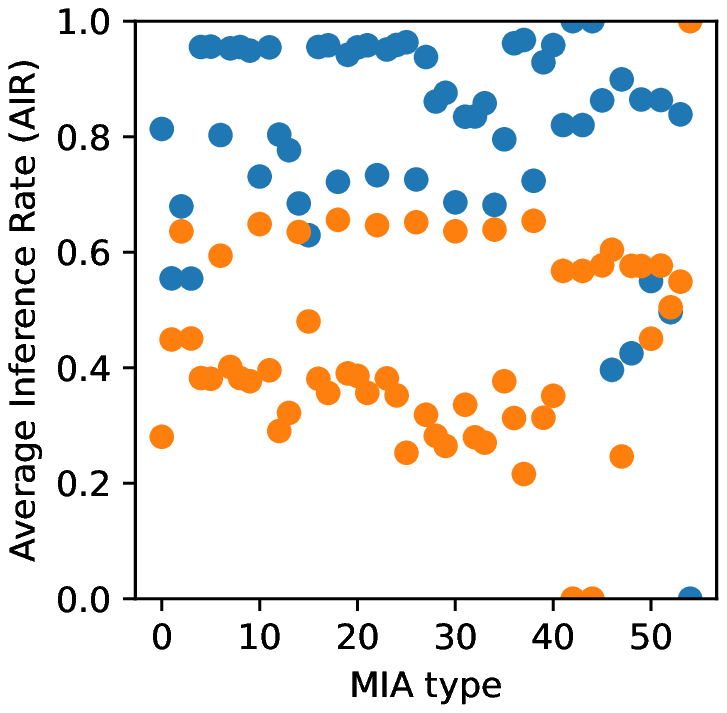}
         \caption{CIFAR-10, ResNet18, Split\_num: 20, Rates: (0.5, 0.5, 0.5)}
     \end{subfigure}
     \begin{subfigure}[b]{0.3\textwidth}
         \centering
         \includegraphics[scale=0.6]{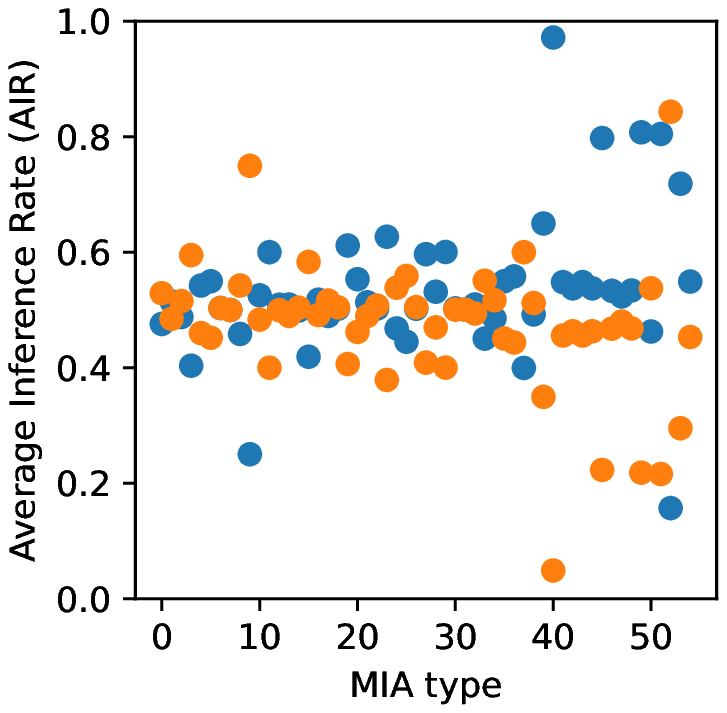}
         \caption{MNIST, CNN, Split\_num: 20, Rates: (0.5, 0.5, 0.5)}
     \end{subfigure}
     \hfill
     \begin{subfigure}[b]{0.3\textwidth}
         \centering
         \includegraphics[scale=0.6]{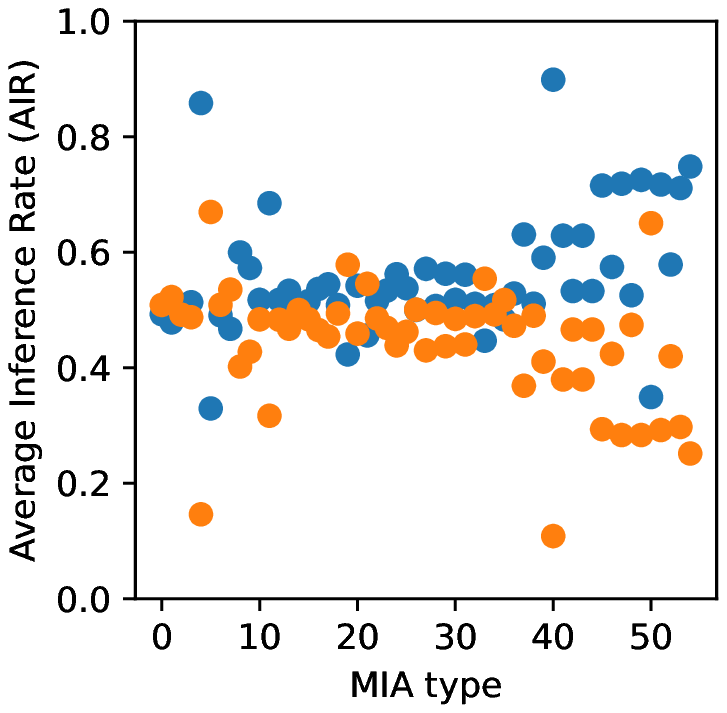}
         \caption{MNIST, CNN, Split\_num: 40, Rates: (0.5, 0.5, 0.5)}
     \end{subfigure}
     \hfill
     \begin{subfigure}[b]{0.3\textwidth}
         \centering
         \includegraphics[scale=0.6]{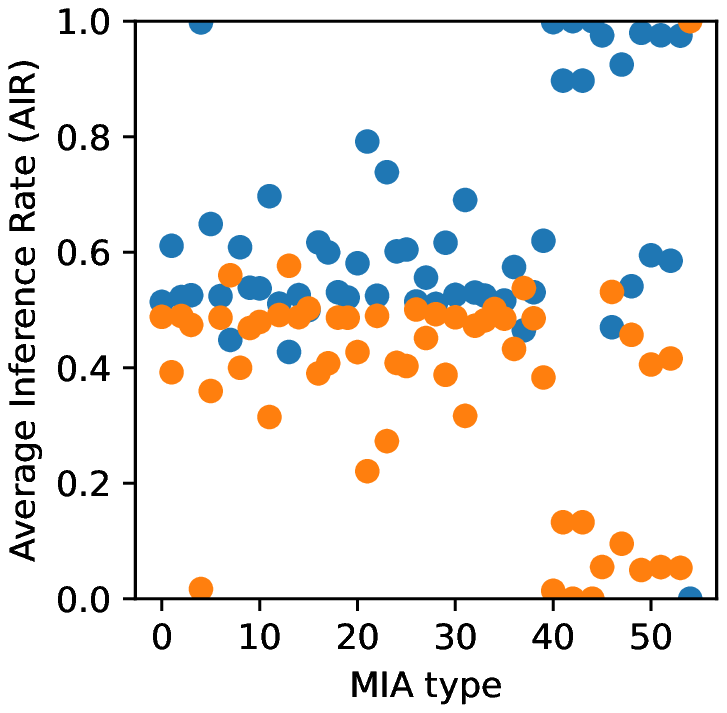}
         \caption{MNIST, ResNet18, Split\_num: 20, Rates: (0.5, 0.5, 0.5)}
     \end{subfigure}
     \hfill
     \begin{subfigure}[b]{0.3\textwidth}
         \centering
         \includegraphics[scale=0.6]{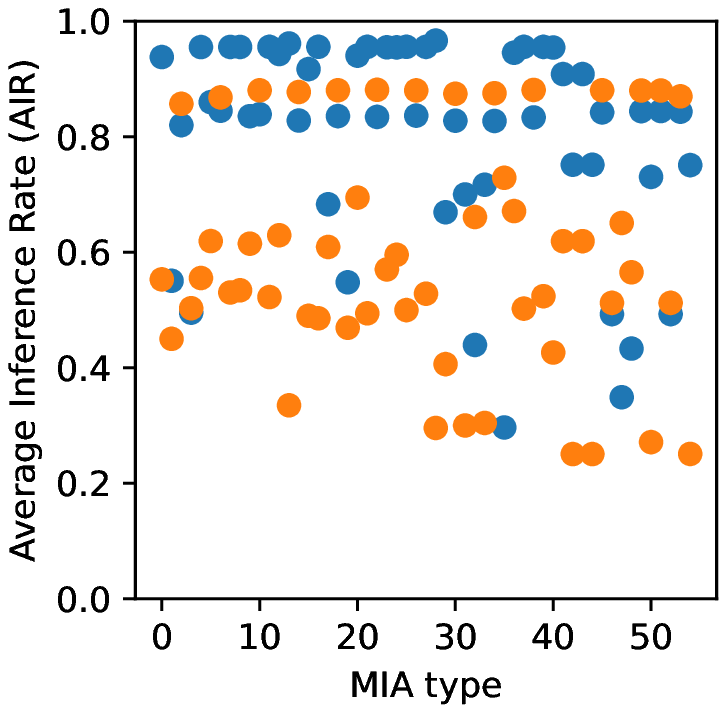}
         \caption{PURCHASE-100, Shallow MLP, Split\_num: 20, Rates: (0.5, 0.5, 0.5)}
     \end{subfigure}
     \hfill
     \begin{subfigure}[b]{0.3\textwidth}
         \centering
         \includegraphics[scale=0.6]{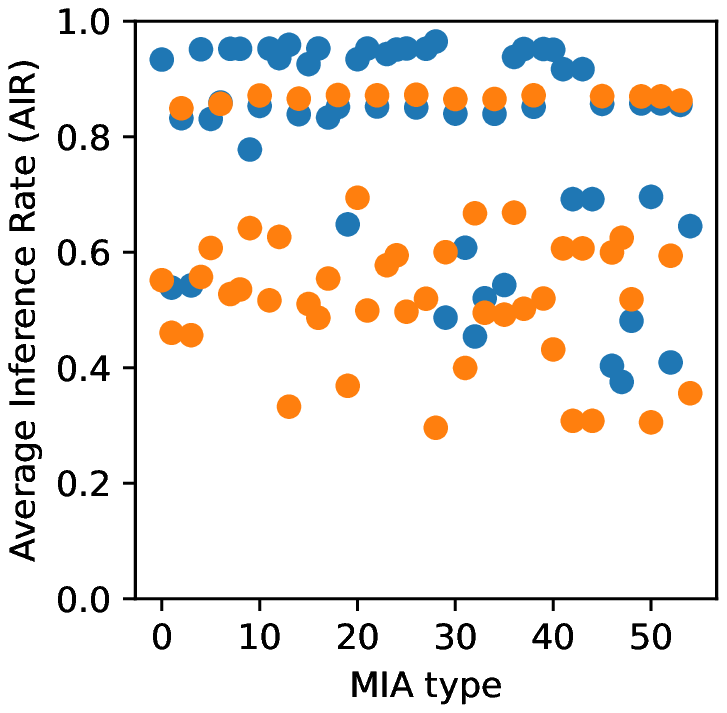}
         \caption{PURCHASE-100, Shallow MLP, Split\_num: 40, Rates: (0.5, 0.5, 0.5)}
     \end{subfigure}
     \hfill
     \begin{subfigure}[b]{0.3\textwidth}
         \centering
         \includegraphics[scale=0.6]{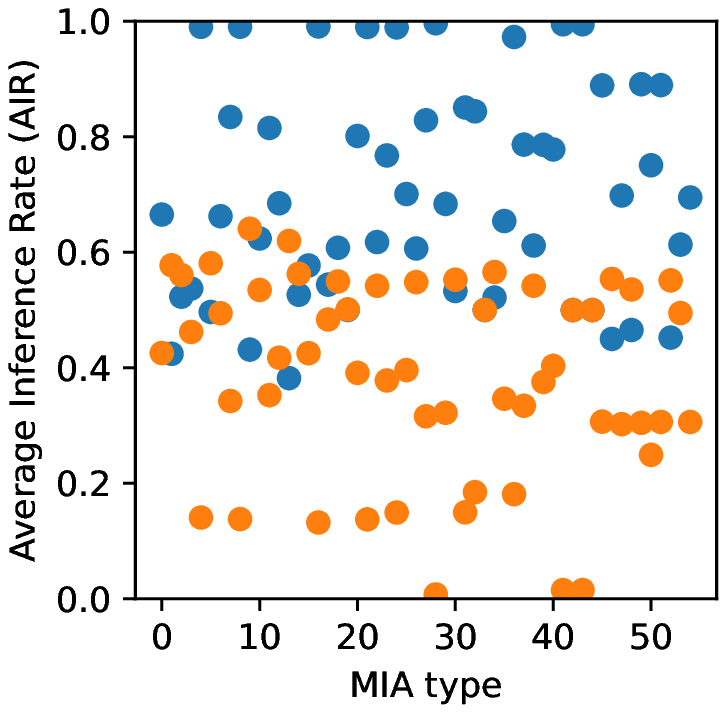}
         \caption{PURCHASE-100, Deeper MLP, Split\_num: 20, Rates: (0.5, 0.5, 0.5)}
     \end{subfigure}
        \caption{The AMIR and ANMIR values of MIAs under different configurations. The definitions of the AMIR and ANMIR metrics are in equations~\eqref{AMIR_formula} and~\eqref{ANMIR_formula}. The x-axis means the type of MIA (54 kinds of MIAs). The y-axis represents the specific value of metrics for each MIA.}
        \label{Average_inference_rate_of_different_MIAs}
\end{figure*}

Apart from the data point's exposure rate, we explore MIA's inference rate, which represents the inference correctness from the angle of the MIA. From equations~\eqref{MIR_formula},~\eqref{AMIR_formula},~\eqref{NMIR_formula}, and~\eqref{ANMIR_formula}, the MIR and NMIR indicate the MIA's inference to a specific data point. The AMIR and ANMIR show the MIA's inference to all training and testing data points. We expose the AMIR and ANMIR under varying configurations in Figure~\ref{Average_inference_rate_of_different_MIAs}. From the figure, we can observe that the AMIR value is generally larger than the ANMIR value. The blue points in the figure are higher than the red points. For example, in the third sub-figure of CIFAR-10 and three sub-figures of PURCHASE-100, the AMIR is close to 1. This observation is similar to the previous result: the AMER value is higher than the ANMER value. The reason is that training data points are more vulnerable than testing data points, even under one MIA.

Table~\ref{Data_points_with_high_MIR_orNMIR} displays a small number of data points with high MIR or NMIR values while the corresponding MIAs have low AMIR or ANMIR values. In the table, the fourth column shows data points' top ten MIR (NMIR). Those ten data points' MT (NMT) values are exposed in the fifth column. We can observe that the top ten MIR and NMIR values of data points are all 1, with high MT and NMT values from 8 to 13. The high MT and NMT values indicate that MIR and NMIR metrics are not calculated occasionally. This observation suggests MIAs with low AMIR or ANMIR values (close to 0.5) can also infer part of data points with high probability. The reason is that some MIAs have an inference tendency to a piece of data points even though the overall inference performance is relatively low.

\begin{table*}[ht!]
\centering
\caption{Top 10 data points with a high MIR or NMIR while the AMIR and ANMIR are relatively low.}
\label{Data_points_with_high_MIR_orNMIR}
\begin{adjustbox}{max width=1\textwidth}
\begin{tabular}{cc|c|c|c|}
\hline
\multicolumn{1}{|c|}{Configuration}              & MIA type (mark) & AMIR  & Top 10 MIR of data points                                                                         & MT for those 10 data points                                                             \\ \hline
\multicolumn{1}{|c|}{CIFAR-10 (LeNet, Split\_num: 20, Rates: (0.5, 0.5, 0.5))}    & 53              & 0.56  & \begin{tabular}[c]{@{}c@{}}{[}1.0, 1.0, 1.0, 1.0, 1.0, \\ 1.0, 1.0, 1.0, 1.0, 1.0{]}\end{tabular} & \begin{tabular}[|c|]{@{}c@{}}{[}13, 12, 12, 11, 11, \\ 11, 11, 11, 11, 11{]}\end{tabular} \\ \hline
\multicolumn{1}{|c|}{MNIST (CNN, Split\_num: 20, Rates: (0.5, 0.5, 0.5))}    & 1               & 0.51  & \begin{tabular}[c]{@{}c@{}}{[}1.0, 1.0, 1.0, 1.0, 1.0, \\ 1.0, 1.0, 1.0, 1.0, 1.0{]}\end{tabular} & \begin{tabular}[c]{@{}c@{}}{[}10, 9, 8, 8, 8, \\ 8, 8, 8, 8, 8{]}\end{tabular}          \\ \hline
\multicolumn{1}{|c|}{PURCHASE-100 (Shallow MLP, Split\_num: 20, Rates: (0.5, 0.5, 0.5))} & 45              & 0.84  & \begin{tabular}[c]{@{}c@{}}{[}1.0, 1.0, 1.0, 1.0, 1.0, \\ 1.0, 1.0, 1.0, 1.0, 1.0{]}\end{tabular} & \begin{tabular}[c]{@{}c@{}}{[}12, 12, 11, 11, 11, \\ 11, 11, 11, 11, 11{]}\end{tabular} \\\hline \\\hline
\multicolumn{1}{|c|}{Configuration}               & MIA type (mark)       & ANMIR & Top 10 NMIR of data points                                                                        & NMT for those 10 data points                                                            \\ \hline
\multicolumn{1}{|c|}{CIFAR-10 (LeNet, Split\_num: 20, Rates: (0.5, 0.5, 0.5))}    & 52              & 0.51  & \begin{tabular}[c]{@{}c@{}}{[}1.0, 1.0, 1.0, 1.0, 1.0, \\ 1.0, 1.0, 1.0, 1.0, 1.0{]}\end{tabular} & \begin{tabular}[c]{@{}c@{}}{[}12, 12, 11, 11, 11, \\ 11, 10, 10, 10, 10{]}\end{tabular} \\ \hline
\multicolumn{1}{|c|}{MNIST (CNN, Split\_num: 20, Rates: (0.5, 0.5, 0.5))}    & 21              & 0.49  & \begin{tabular}[c]{@{}c@{}}{[}1.0, 1.0, 1.0, 1.0, 1.0, \\ 1.0, 1.0, 1.0, 1.0, 1.0{]}\end{tabular} & \begin{tabular}[c]{@{}c@{}}{[}8, 8, 8, 8, 8, \\ 8, 8, 8, 8, 8{]}\end{tabular}           \\ \hline
\multicolumn{1}{|c|}{PURCHASE-100 (Shallow MLP, Split\_num: 20, Rates: (0.5, 0.5, 0.5))} & 48              & 0.57  & \begin{tabular}[c]{@{}c@{}}{[}1.0, 1.0, 1.0, 1.0, 1.0, \\ 1.0, 1.0, 1.0, 1.0, 1.0{]}\end{tabular} & \begin{tabular}[c]{@{}c@{}}{[}12, 12, 12, 11, 11, \\ 11, 11, 11, 11, 11{]}\end{tabular} \\ \hline
\end{tabular}
\end{adjustbox}
\end{table*}




\subsection{Vulnerable Data Points Comparison}
\label{vulnerable data points}

\begin{figure*}
     \centering
     \begin{subfigure}[b]{0.33\textwidth}
         \centering
         \includegraphics[width=\textwidth]{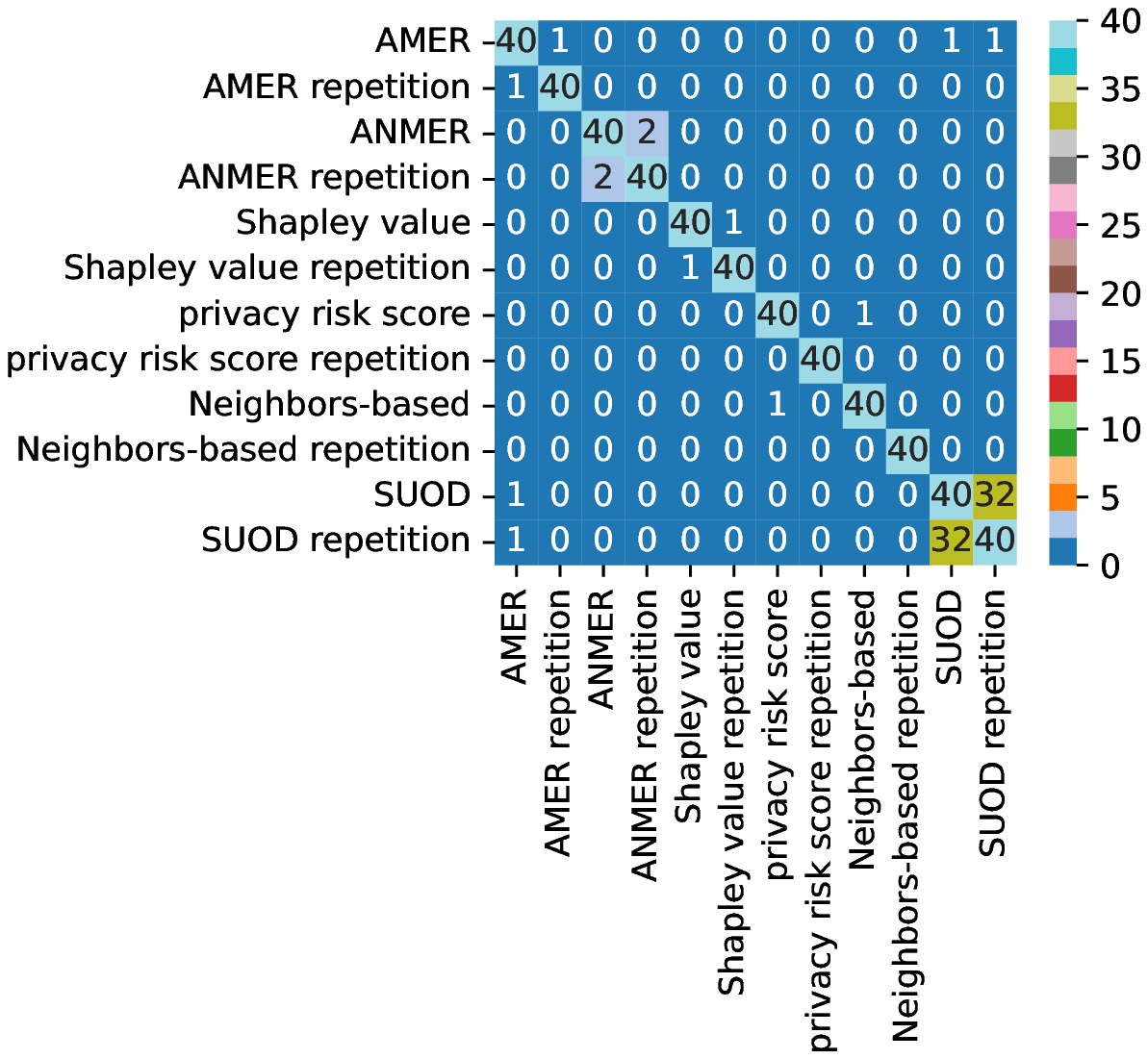}
         \caption{CIFAR-10}
     \end{subfigure}
     \hfill
     \begin{subfigure}[b]{0.33\textwidth}
         \centering
         \includegraphics[width=\textwidth]{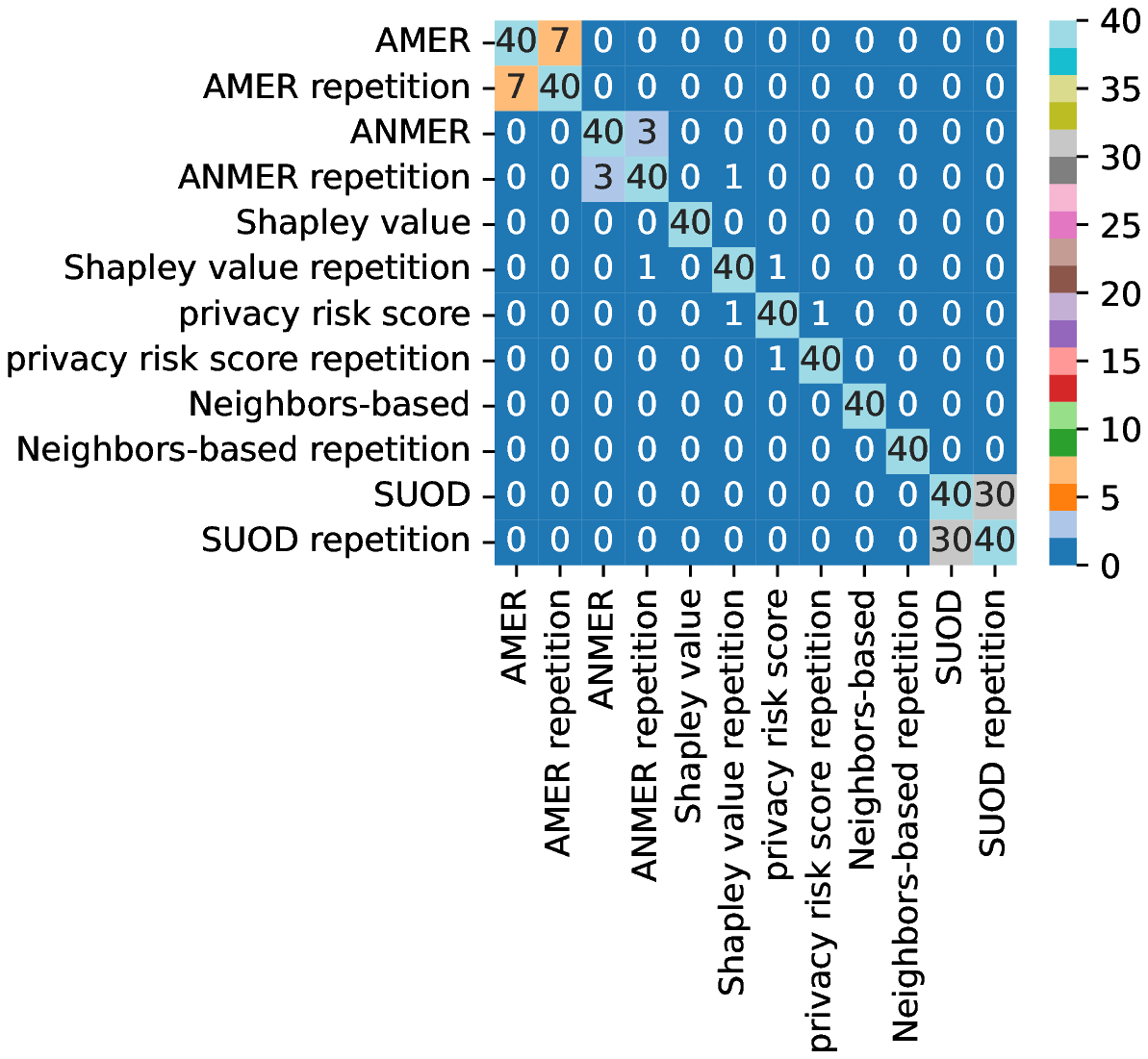}
         \caption{MNIST}
     \end{subfigure}
     \hfill
     \begin{subfigure}[b]{0.33\textwidth}
         \centering
         \includegraphics[width=\textwidth]{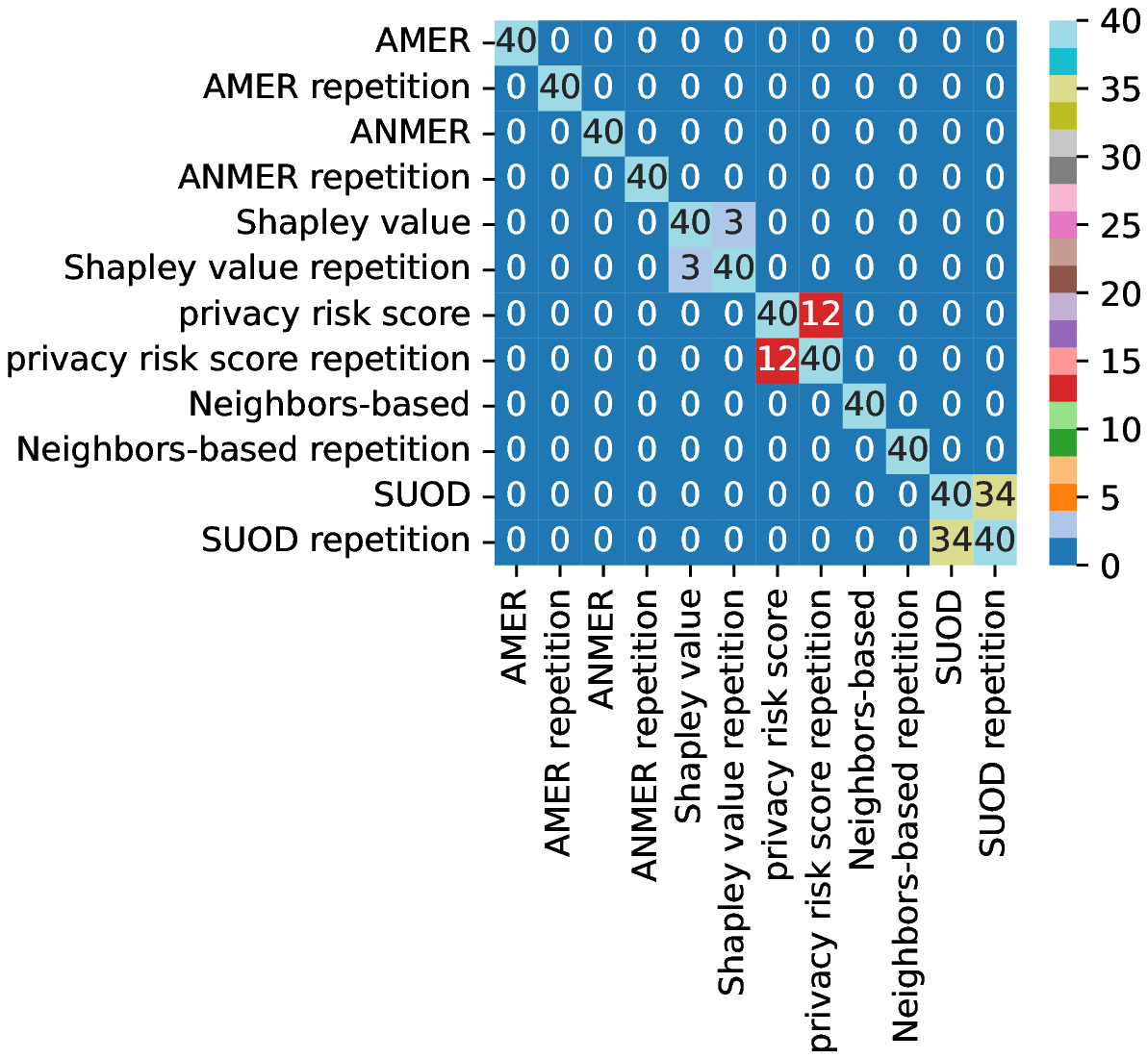}
         \caption{PURCHASE-100}
     \end{subfigure}
        \caption{Overlapping vulnerable data points among different methods within three datasets are exposed. The x-axis and y-axis represent approaches. Tag 'repetition' indicates repeating the experiment without changing anything. The digit within the figure means the number of overlapping data points between two methods represented by the x-axis and y-axis. Settings (Structure: LeNet (CIFAR-10), CNN (MNIST), Shallow MLP (PURCHASE-100), Split\_num: 20 for three datasets, Rates: (0.5, 0.5, 0.5) for three datasets) are the same while finding vulnerable data points on the same dataset with methods except for the Neighbors-based approach and SUOD (Section~\ref{find_and_compare_vulnerable_data_points}), which attribute to the specific steps of those two manners.}
        \label{overlapping_v_data_points}
\end{figure*}

We determine 40 vulnerable data points under multiple MIAs and target models based on our new metrics (AMER and ANMER), neighbors-based method, privacy risk score, and Shapley value. Besides, the outlier detection method, SUOD, is selected for comparison with previous methods. The data points found by the SUOD are not proven by previous work as truly vulnerable data points. Figure~\ref{overlapping_v_data_points} displays the overlapping data points between varying vulnerable sets.

  

There are almost no overlapping data points among different vulnerable data points detection methods. Previous methods do not find the vulnerable data points with our new metrics except for a few data points. As the previous analysis, our new metrics can reflect and capture vulnerable data points under multiple MIAs and target models. Therefore, previous methods are unsuitable for detecting vulnerable data points under multiple MIAs and target models. In other words, previous methods determine the vulnerable data points for a specific MIA and target model are not vulnerable data points under multiple MIAs and target models. In addition, we cannot obtain similar data points apart from the SUOD method by repeating the same experiment without changing anything. The target model and MIA are slightly different after repetition, even though the dataset, model structure, training process, and hyperparameters are identical. However, the vulnerable data points under multiple MIAs and target models differ after repetition. From those observations and analyses, we conclude that vulnerability is not the fixed characteristic of the data point. Oppositely, it is related to the target model and MIA. Therefore, vulnerable data points are distinctive while repeating the same experiment. Finally, the vulnerable data points determined by the AMER and ANMER values are different except for a few data points. This observation comes from the small number of overlapping vulnerable data points determined by the AMER and ANMER values. The reason is that the vulnerability of the data point is different while this data point is in the training and testing data. Therefore, we cannot find similar vulnerable data points with AMER and ANMER values.


\section{Related Work}

After the seminal work of Shokri et al.~\cite{shokri_membership_2017}, Membership Inference Attack has attracted significant attention. Current research works are mainly divided into attacking methods, defense strategies, and a deeper understanding of the MIA. 

\textbf{Attacking methods.} Shokri et al. put forward the application of the shadow model to mimic the behavior of the target model and obtain the dataset for the attack model training~\cite{shokri_membership_2017}. Salem et al. relaxed some assumptions in the work of Shokri et al.. They proposed a data transferring attack and threshold-based attacks using the highest posterior, standard deviation, and entropy.~\cite{salem_ml-leaks_2018}. 
Yeom et al. formulated three adversaries of the MIAs with additional information, including loss, empirical error, leave-one-out validation error, average training loss, and even the training dataset~\cite{yeom_privacy_2018}. Sablayrolles et al. used Bayes' formula, optimal steps, and approximations to implement optimal MIA only with loss~\cite{sablayrolles_white-box_2019}. Long et al. carefully selected a few vulnerable data points and attacked them with direct and indirect inference~\cite{long_pragmatic_2020}. Li et al.~\cite{li_label-leaks_2020} and Choquette-choo et al.~\cite{choquette-choo_label-only_2021} investigated attacking target model under label-only condition. They proposed three methods: applying adversarial examples, data augmentation, and relabelling shadow data with the target model. Carlini et al. presented the Likelihood Ratio Attack (LiRA), which formulates the MIA as hypothesis testing and considers the data point in or out of the training data separately~\cite{carlini_membership_2021}. 
The manipulation of separation inspires our work. Our empirical result shows that the data point's vulnerability is different in the training data compared with the testing data. Ye et al. put forward model-dependent and sample-dependent MIA via distillation, which means the threshold of determination is related to the target model and data point~\cite{ye_enhanced_2021}.

\textbf{Defense strategies.} Strategies for reducing the overfitting are proposed for eliminating the MIA, including dropout~\cite{srivastava_dropout_2014}, L2-norm standard regularization,  and model stacking. Some strategies manipulate the target model's output, including classes of output limitation, prediction vector modification, and prediction entropy increase~\cite{shokri_membership_2017,salem_ml-leaks_2018}. Nasr et al. combined the training process of the target model with a misleading attack classifier, decreasing the performance of classifier-based MIAs~\cite{nasr_machine_2018}. Jia et al. added slight perturbation to the prediction vector, leading to the misclassification of classifier-based MIAs~\cite{jia_memguard_2019}. Shejwalkar et al. leveraged knowledge distillation to train ML models with membership privacy~\cite{shejwalkar_membership_2019}. Li et al. utilized the mix-up data augmentation and Maximum Mean Discrepancy regularization to close training and validation accuracy, decreasing the attack performance~\cite{li_membership_2021}. Tang et al. proposed a novel ensemble architecture and a self-distillation framework to defeat MIAs~\cite{tang_novel_2021}. Jarin et al. decreased the performance of MIAs by excluding the prediction of sub-model whose training data include the target data point~\cite{jarin_miashield_2022}. Finally, DP-SGD is frequently mentioned to defend against MIAs with high utility costs~\cite{abadi_deep_2016}.

\textbf{Deeper understanding.} The factors contributing to the success of the MIA attracted frequent discussion. Overfitting, the choice of target model and dataset, the selection of part data points, and the complexity of the training dataset are recognized as influence factors of MIAs~\cite{shokri_membership_2017,salem_ml-leaks_2018,long_pragmatic_2020,yaghini_disparate_2019,yeom_privacy_2018,truex_towards_2018,liu_ml-doctor_2021}. Besides, some works are devoted to analyzing the privacy risk of the target model with the help of the MIA~\cite{murakonda_ml_2020,liu_ml-doctor_2021}. Furthermore, the data points with high privacy risk, susceptibility, or vulnerability are detected with different methods~\cite{song_systematic_2021,duddu_shapr_2021,long_pragmatic_2020}. The focus of this paper, vulnerable data points under multiple MIAs and target models, is different from previous settings.

\section{Limitations}

There are two main limitations to our work. First, we only consider classification tasks with the image and number features. The MIAs to other tasks and data formats are not included, e.g., generative models~\cite{chen_gan-leaks_2020,hu_membership_2021,hayes_logan_2019}, graph data~\cite{wu_adapting_2021,he_node-level_2021,olatunji_membership_2021,zhang_inference_2021}, and federated leaning~\cite{melis_exploiting_2018,nasr_comprehensive_2019}.

Second, the type of MIAs implemented in current research is limited. MIAs in this work have to use the same distribution's shadow dataset to determine the threshold or train the attack model. There are some strategies to overcome the condition that the shadow dataset comes from same distribution~\cite{salem_ml-leaks_2018,salem_ml-leaks_2018,li_label-leaks_2020}. Besides, some MIAs proposed recently are not implemented in this work. For example, the label-only MIAs proposed by Li et al.~\cite{li_label-leaks_2020} and Choquette-choo et al.~\cite{choquette-choo_label-only_2021} with the help of adversarial examples or data augmentation, the LiRA put forward by Carlini et al.~\cite{carlini_membership_2021}, and the strategy of calculating threshold for each category proposed by Song et al.~\cite{song_systematic_2021}.
While our work has those two limitations (which we leave for future work), we can still conclude results unrelated to specific MIAs and datasets.


\section{Conclusions}
This paper explores a single data point's vulnerability and tries to find vulnerable data points under multiple MIAs and target models. To formally analyze data points' vulnerability, we define metrics about data points' exposure rate and MIA's inference rate. All experiments are completed with the help of our newly developed platform, VMIAP, which is scalable and flexible for attacking target models with varying MIAs. 

From the analysis of MIAs' attacking performance, the accuracy gap, measured by the gap between the training and testing accuracy, influences the attack accuracy. Generally, a large accuracy gap contributes to high attack accuracy. Furthermore, the dataset and model structure are non-negligible factors of MIAs' attack performance. It gives us the inspiration that we need to select suitable MIA for a specific combination of dataset and model structure to get high attack performance. In addition, our new metrics, AMER and ANMER, can reflect the actual situation of data points' vulnerability and capture vulnerable data points under multiple MIAs and target models. We observe that the AMER values are higher than the ANMER values. It means that if the data point appears in the training data of target models, its vulnerability is usually higher than in the testing data. Moreover, MIA has an inference tendency to some data points even though its overall inference performance is relatively low. Our results indicate that the vulnerability under multiple MIAs and target models is not the fixed characteristic of data points. On the contrary, it is related to MIAs and target models. Finally, we find neighbors, privacy risk score, and Shapley value-based methods are unsuitable for detecting vulnerable data points under multiple MIAs and target models.


%



\section*{Acknowledgments}
This research is supported by the Chinese Scholarship Council (CSC).


\ifCLASSOPTIONcaptionsoff
  \newpage
\fi



%
\bibliographystyle{IEEEtran}
\bibliography{paper_citation}




%

\begin{IEEEbiography}[{\includegraphics[width=1in,height=1.25in,clip,keepaspectratio]{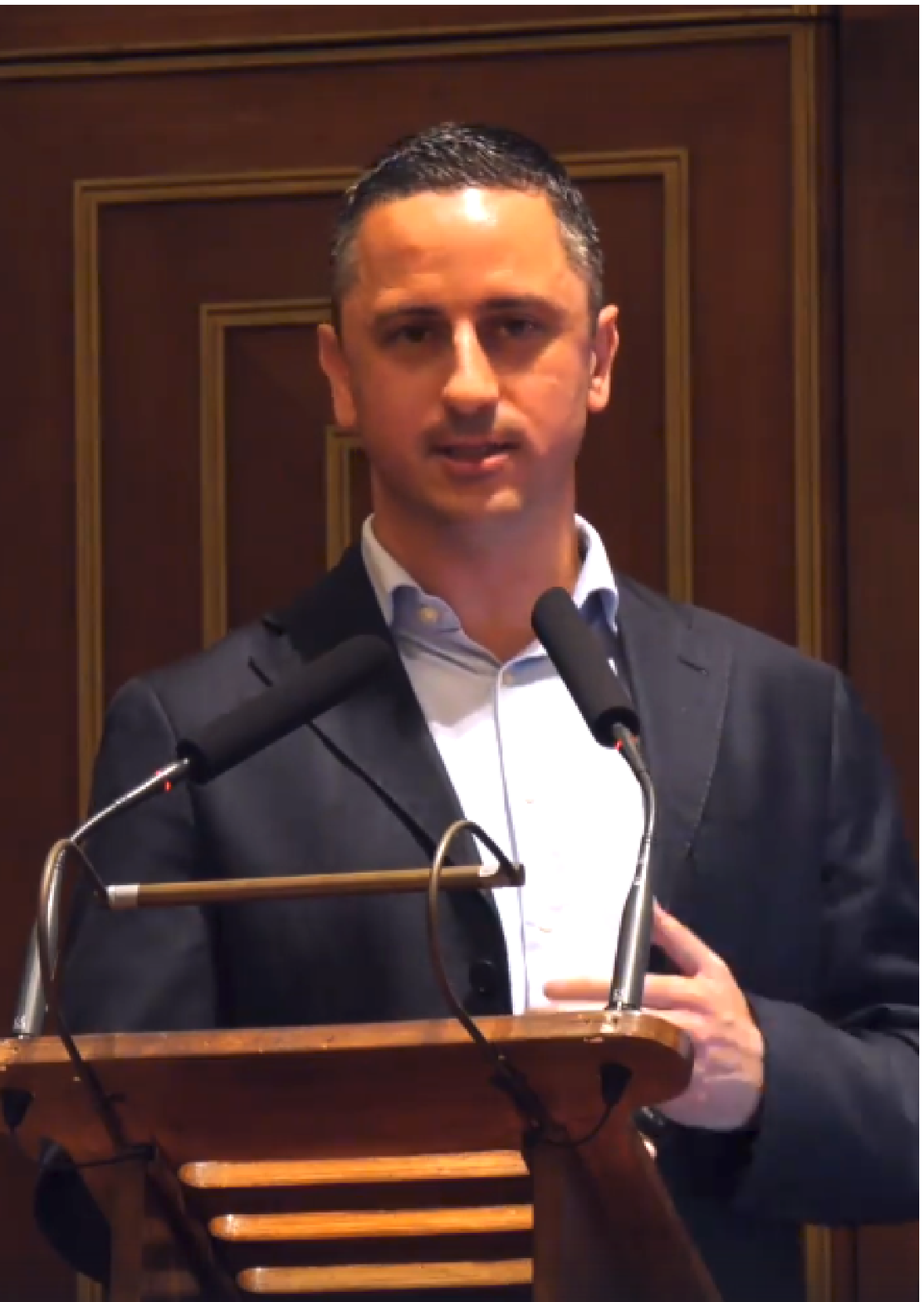}}]{Mauro Conti} is Full Professor at the University of Padua, Italy. He is also affiliated with TU Delft and the University of Washington, Seattle. He obtained his Ph.D. from Sapienza University of Rome, Italy, in 2009. After his Ph.D., he was a Post-Doc Researcher at Vrije Universiteit Amsterdam, The Netherlands. In 2011 he joined as Assistant Professor at the University of Padua, where he became Associate Professor in 2015, and Full Professor in 2018. He has been Visiting Researcher at GMU, UCLA, UCI, TU Darmstadt, UF, and FIU. He has been awarded with a Marie Curie Fellowship (2012) by the European Commission, and with a Fellowship by the German DAAD (2013). His research is also funded by companies, including Cisco, Intel, and Huawei. His main research interest is in the area of Security and Privacy. In this area, he published more than 450 papers in topmost international peer-reviewed journals and conferences. He is Editor-in-Chief for IEEE Transactions on Information Forensics and Security, Area Editor-in-Chief for IEEE Communications Surveys \& Tutorials, and has been Associate Editor for several journals, including IEEE Communications Surveys \& Tutorials, IEEE Transactions on Dependable and Secure Computing, IEEE Transactions on Information Forensics and Security, and IEEE Transactions on Network and Service Management. He was Program Chair for TRUST 2015, ICISS 2016, WiSec 2017, ACNS 2020, CANS 2021, and General Chair for SecureComm 2012, SACMAT 2013, NSS 2021 and ACNS 2022. He is a Fellow of the IEEE, a Senior Member of the ACM, and a Fellow of the Young Academy of Europe.
\end{IEEEbiography}


\vspace{-6.2cm}
\begin{IEEEbiography}[{\includegraphics[width=1in,height=1.25in,clip,keepaspectratio]{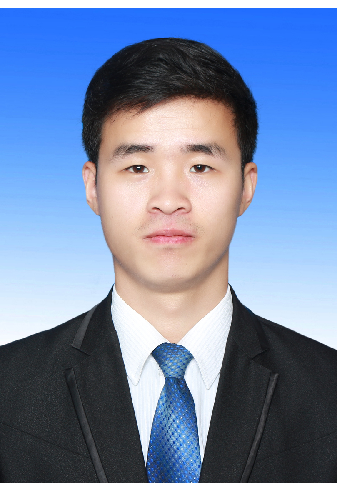}}]{Jiaxin Li}
is a PhD student at the University of Padua, Italy. He focuses on the research of machine learning security, including membership inference attacks, adversarial examples, and differential privacy. He obtains his bachelor's and master's degrees from the Harbin Institute of Technology, China.
\end{IEEEbiography}

\vspace{-6.2cm}
\begin{IEEEbiography}[{\includegraphics[width=1in,height=1.25in,clip,keepaspectratio]{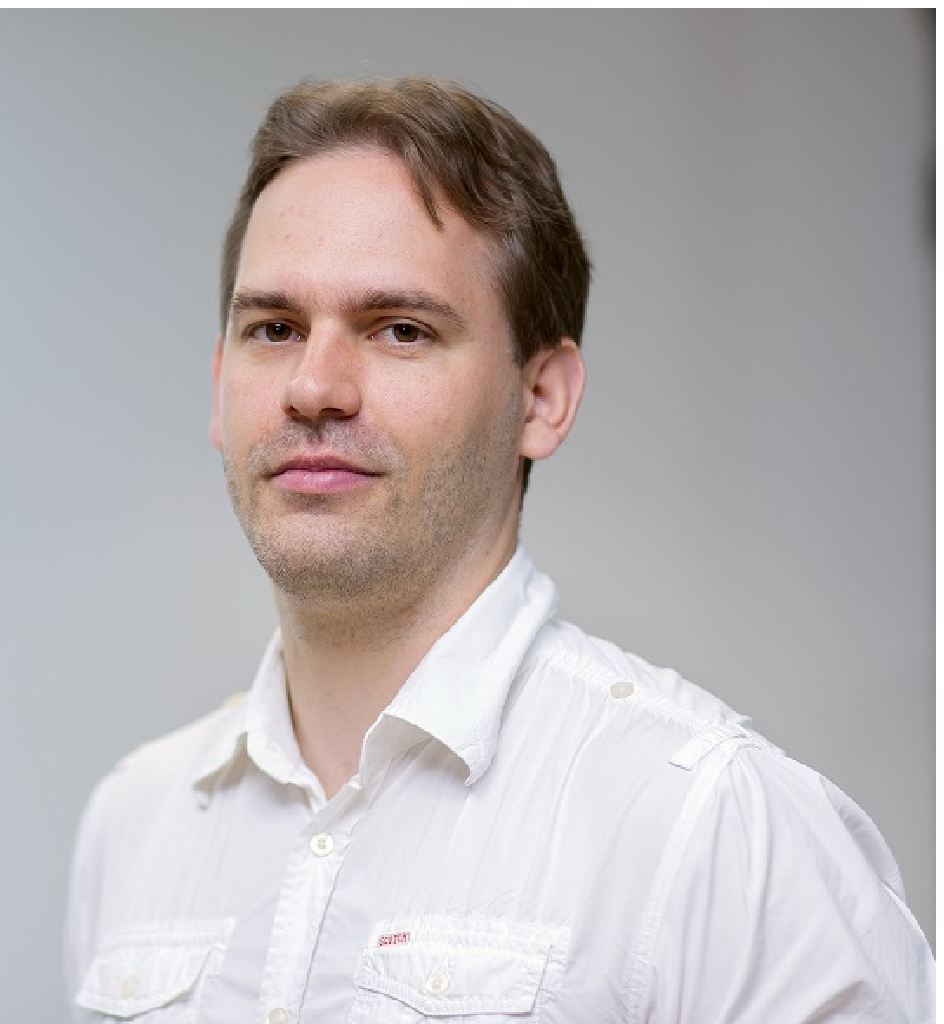}}]{Stjepan Picek}
is an associate professor at the Radboud University, The Netherlands. 
His research interests include security, machine learning, and evolutionary algorithms. Before Radboud, Stjepan was assistant professor at Delft University of Technology and a postdoc at MIT, USA and KU Leuven, Belgium.
\end{IEEEbiography}




\end{document}